\documentclass[preprint]{aastex} 
\usepackage{natbib}
\newcommand\lae{\mathrel{<\kern-1.0em\lower0.9ex\hbox{$\sim$}}}
\newcommand\gae{\mathrel{>\kern-1.0em\lower0.9ex\hbox{$\sim$}}}
\newcommand\kms{km~s$^{-1}$}
\newcommand\mone{$^{-1}$}
\newcommand\mtwo{$^{-2}$}

\newcommand{\m}{$\mu$m}

\newcommand{\molhyd}{H$_2$}

\title{Infrared Spectral Energy Distributions of Seyfert Galaxies:
  Spitzer Space Telescope Observations of the 12 \m\ Sample of Active
  Galaxies} 
\begin{document}

\author{J. F. Gallimore\altaffilmark{1,2}, A. Yzaguirre\altaffilmark{1,3}, 
  J. Jakoboski\altaffilmark{1,4}, M. J. Stevenosky\altaffilmark{1,5},
  D. J. Axon\altaffilmark{6,7}, S. A. Baum\altaffilmark{8}, 
  C. L. Buchanan\altaffilmark{9}, M. Elitzur\altaffilmark{10},
  M. Elvis\altaffilmark{11}, C. P. O'Dea\altaffilmark{6},
  A. Robinson\altaffilmark{6}}

\author{draft:\today}

\altaffiltext{1}{Department of Physics and Astronomy, Bucknell
  University, Lewisburg, PA 17837} 
\altaffiltext{2}{Currently on leave at NRAO, 520 Edgemont Rd.,
  Charlottesville, VA 22903}
\altaffiltext{3}{Department of Physics, California State University,
  Fullerton, P.O. Box 6866, Fullerton, CA 92834-6866}
\altaffiltext{4}{Jet Propulsion Laboratory, 4800 Oak Grove Drive,
  Pasadena, CA 91190}
\altaffiltext{5}{Franklin Pierce Law Center, Two White Street,
  Concord, NH 03301}
\altaffiltext{6}{Department of Physics, Rochester Institute of
  Technology, 84 Lomb Memorial Drive, Rochester, NY 14623}
\altaffiltext{7}{School of Mathematical and Physical Sciences,
  University of Sussex, Falmer, Brighton, BN1 9QH, UK}
\altaffiltext{8}{Chester F. Carlson Center for Imaging Science,
  Rochester Institute of Technology, 54 Lomb Memorial Drive,
  Rochester, NY 14623} 
\altaffiltext{9}{School of Physics, University of Melbourne,
  Parkville, Victoria, 3010 Australia} 
\altaffiltext{10}{Department of Physics and Astronomy, University of
  Kentucky, Lexington, KY 40506} 
\altaffiltext{11}{Harvard-Smithsonian Center for Astrophysics, 60
  Garden Street,  Cambridge, MA 02138}

\begin{abstract}

The mid-infrared spectral energy distributions (SEDs) of 83 active
galaxies, mostly Seyfert galaxies, selected from the extended 12
\m\ sample are presented. The data were collected using all three
instruments, IRAC, IRS, and MIPS, aboard the {\em Spitzer Space
  Telescope}. The IRS data were obtained in spectral mapping mode, and
the photometric data from IRAC and IRS were extracted from matched,
20\arcsec\ diameter circular apertures. The MIPS data were obtained in
SED mode, providing very low resolution spectroscopy ($R \sim 20$)
between $\sim 55$ and 90 \m\ in a larger, 20\arcsec\ $\times$
30\arcsec\ synthetic aperture. We further present the data from a
spectral decomposition of the SEDs, including equivalent widths and
fluxes of key emission lines; silicate 10~\m\ and 18~\m\ emission and
absorption strengths; IRAC magnitudes; and mid-far infrared spectral
indices. Finally, we examine the SEDs averaged within optical
classifications of activity. We find that the infrared SEDs of Seyfert
1s and Seyfert 2s with hidden broad line regions (HBLR, as revealed by
spectropolarimetry or other technique) are qualitatively similar,
except that Seyfert 1s show silicate emission and HBLR Seyfert 2s show
silicate absorption. The infrared SEDs of other classes within the
12~\m\ sample, including Seyfert 1.8-1.9, non-HBLR Seyfert 2 (not yet
shown to hide a type 1 nucleus), LINER, and HII galaxies, appear to be
dominated by star-formation, as evidenced by blue IRAC colors, strong
PAH emission, and strong far-infrared continuum emission, measured
relative to mid-infrared continuum emission.
\end{abstract}


\section{Introduction}

Three components dominate the mid-infrared spectrum of an active
galaxy \citep{2000ARA&A..38..761G}: (1) thermal radiation from a
dusty, compact medium that surrounds the active nucleus (AGN) and can
obscure direct sight-lines to it; (2) PAH features and thermal dust
continuum associated with star-formation or perhaps a powerful
starburst; and (3) line features arising from molecular, atomic, and
ionic species.

The dusty medium surrounding the AGN is commonly referred to as ``the
dusty torus.'' Its presence is inferred for many AGNs based, for
example, on spectropolarimetry \citep{1985ApJ...297..621A,
  1993ARA&A..31..473A, 1995PASP..107..803U, 2004MNRAS.350..140S},
X-ray spectroscopy \citep{2007ApJ...659L.111R, 2002ApJ...571..234R},
and infrared aperture synthesis measurements
\citep{2004Natur.429...47J, 2007A&A...474..837T}. Observations further
indicate that the dusty medium must be axisymmetric, permitting low
extinction sight-lines to type 1 AGNs, where the broad-line region
(BLR) is apparent in total intensity (Stokes $I$) spectra (the
``pole-on'' view), but high-extinction sight-lines to type 2 AGNs,
which show suppressed broad line emission and AGN continuum (the
``edge-on'' view).

Since the dusty torus re-radiates incident AGN continuum in the
mid-infrared, its SED provides an indirect measure of the AGN
luminosity \citep{2008ApJ...685..160N}, important especially for
heavily obscured AGNs where other indirect diagnostics may not be
available. Mid-infrared fine-structure lines can also constrain the
intrinsic shape of the AGN SED, because particular line ratios are
sensitive to the shape of the SED but less sensitive to extinction
compared to optical/UV lines of species with similar ionization
energies \citep{2000ApJ...536..710A}.

The torus SED depends on the geometry and clumpiness of the torus,
among other properties \citep{2002ApJ...570L...9N,
  2008ApJ...685..160N}. For example, smooth, cylindrical tori produce
very strong 10 \m\ silicate (Sil) features, whether in emission when
viewed more nearly pole-on, or deep absorption when viewed edge-on
\citep{1992ApJ...401...99P}. Clumpy tori, somewhat independent of the
assumed geometry, instead produce much weaker Sil features, because
inter-clump sight-lines can provide a view of hotter dust on the far
side of the torus; our view of clumpy tori includes a mix of cold and
hot clump surfaces that dilutes the Sil features
\citep{2008ApJ...685..160N}. Individual clumps are heated from outside
and cannot produce the strong absorption that a centrally heated dust
shell can have
\citep{2008ApJ...678..729S,2008ApJ...685..147N,2007ApJ...654L..45L}.

The mid-infrared SED is also an important diagnostic of
star-formation. Young star clusters embedded in GMCs are predicted to
produce weaker PAH equivalent width in comparison to older clusters, owing to
photodestruction and continuum dilution by hot dust grains
\citep{2000MNRAS.313..734E}. In global models, PAH features are
sensitive to the fractional luminosity of OB associations and the gas
density of their surrounding ISM \citep{2007A&A...461..445S}. Since
the dusty medium re-radiates incident stellar radiation, the mid-far
infrared luminosity constrains the luminous contribution of
star-formation. 

The mid-infrared SED of active galaxies therefore contains diagnostics
of the AGN, its surrounding dusty torus, and any surrounding
star-formation. The described tracers can, in principle, be
disentangled by spectral decomposition of the global SED
\citep{2003MNRAS.343..585F,2007ApJ...670..129M}. At the present,
available infrared instruments, at least in their default mode of use,
suffer from mismatched apertures leading to discontinuities in the
observed SEDs or mismatched coverage of spectral lines that bias line
ratios.


To remedy this problem of aperture matching, we have used the {\em
  Spitzer Space Telescope} to observe a sample of active galaxies,
  mostly Seyferts, in synthetically matched, 20\arcsec\ diameter
  apertures spanning $\lambda$3.6-36 \m\ and a larger
  aperture, $\sim 20\arcsec \times 30\arcsec$, covering $\lambda$55-90
  \m. These moderate to low resolution SEDs are well-suited for
  spectral decomposition, studies of the PAH and Sil features, as well
  as global constraints on the AGN and star-formation contributions to
  the luminosity.

The observations and data obtained for this survey are presented in
this work. Section~\ref{sec:sample} describes the sample
selection. Section~\ref{sec:observations} provides a detailed account
of the observations and data reduction, with attention to artifact
correction, synthetic aperture matching, and corrections for extended
emission. Section~\ref{sec:analysis} summarizes the extraction of
spectral features using a modified version of the PAHFIT tool
\citep{2007ApJ...656..770S} and measurements based on the
line-subtracted SEDs. We conclude in
Section~\ref{sec:conclusions} with a summary of the properties of the
sample and avenues for future analysis.

\section{Sample Selection}\label{sec:sample}


The sample, listed in Table~\ref{tab:sample}, comprises a subset of
the extended 12 \m\ sample of AGNs \citep{1993ApJS...89....1R}. This
parent sample was defined by an IRAS detection at 12 \m, $F_{12}({\rm
  IRAS}) > 0.22$~Jy, and color selection $F_{60}({\rm IRAS}) > 0.5 F_{12}({\rm
  IRAS})$ or $F_{100}({\rm IRAS}) > F_{12}({\rm IRAS})$ to remove
stars but few galaxies. AGNs were
identified based on prior (usually optical) classification. We
restricted the sample to include (1) only those objects categorized by
\citet{1993ApJS...89....1R} as Seyferts or LINERs, and (2) only those
sources with $cz < 10,000$~\kms. Three sources were removed subsequent
to observations owing either to pointing errors or saturation in the
Spitzer observations: NGC~1068, M-3-34-64, and NGC~4922. Our sample
ultimately includes 83 Seyfert and LINER nuclei.

The main advantage of this sample over other AGN catalogs is its large
collection of published and archival multiwavelength observations
\citep{1996ApJ...473..130R, 1999ApJ...516..660H, 1999ApJ...510..637H,
  2001MNRAS.325..737T, 2000MNRAS.314..573T,
  2006AJ....132..546G,2002ApJ...572..105S}. In addition, there are
comparable numbers of Seyfert 1 (S1) and Seyfert 2 (S2) nuclei, and
their redshift distributions are statistically indistinguishable
\citep{2001MNRAS.325..737T, 2000MNRAS.314..573T}.

\begin{figure}
\plotone{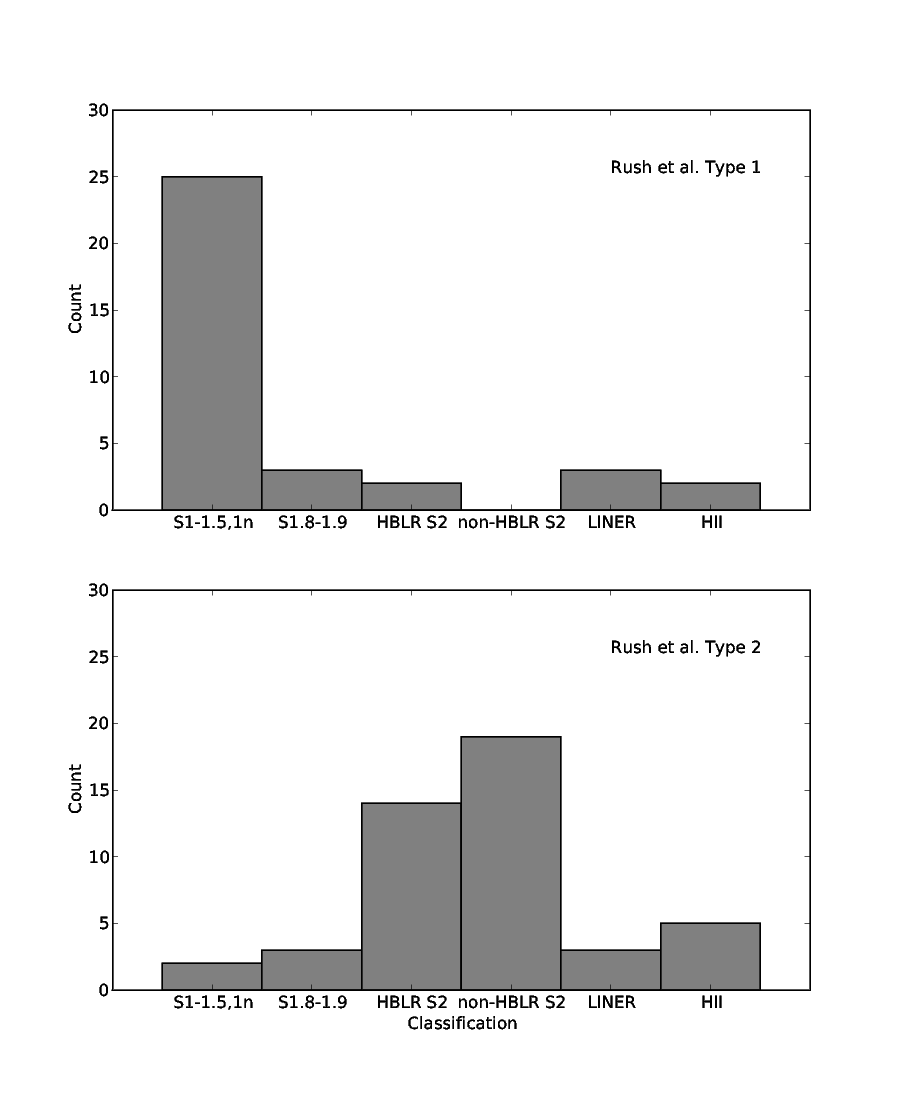}
\caption{A comparison of AGN classifications from \citet{1993ApJS...89....1R} and our
  revised classifications collected from the literature,
  Table~\ref{tab:sample}. The top panel illustrates the
  re-classification of Rush et al. Type~1 AGNs, and the bottom panel
  similarly shows the re-classification of Type~2 AGNs. Seyfert 2s are
split into two groups: (1) HBLR S2s, which are known to harbor an
HBLR, and (2) non-HBLR S2s, in which no HBLR has yet been
detected.}\label{fig:sampleclasses} 
\end{figure}

The original survey paper of \citet{1993ApJS...89....1R} broadly
assigned optical classifications of type 1 (broad-line AGN) and type 2
(narrow-line AGN), but \citet{2003ApJ...583..632T} pointed out that,
even given such coarse binning of activity type, there were many
misclassifications. To aid in a more sophisticated comparison of
infrared properties to optical classifications, we endeavored to
collect updated and more precise classifications from the
literature. The revised classifications are included in
Table~\ref{tab:sample} and illustrated in
Figure~\ref{fig:sampleclasses}. We find that 20\% (7 / 35) of the Rush et
al. Type 1s are re-classified as hidden broad line region (HBLR) Seyfert 2s
(S1h or S1i), LINER, or HII (star-forming
galaxy), and 28\% (13 / 46) of the Rush et al. Type 2s are
re-classified as S1.n, LINER, or HII. Note that HBLRs have been sought
in all but three of the 20 S2s in our survey: NGC~1125, E33-G2, and
NGC~4968 \citep{2003ApJ...583..632T}.

\section{Observations and Data Reduction}\label{sec:observations}

The sample galaxies were observed using all of the instruments of the
{\em Spitzer Space Telescope} (Program ID 3269, Gallimore, P.I.): the
four broadband channels (3.6 \m, 4.5 \m, 5.8 \m, and 8.0 \m) of the
Infrared Array Camera \citep[IRAC; ][]{2004ApJS..154...10F}; the low
resolution gratings of the Infrared Spectrograph 
\citep[IRS; ][]{2004ApJS..154...18H}, operating in spectral mapping mode; and
the Multiband Imaging Photometer for Spitzer 
\citep[MIPS; ][]{2004ApJS..154...25R}, operating in SED mode. The resulting
SEDs are provided in Figure~\ref{fig:allseds}.

Several sample galaxies were observed as part of other {\em Spitzer}
programs that made use of different observing strategies; for example,
in some cases only single-pointing (``Staring Mode'') IRS spectra are
available. We summarize the observations and data reduction techniques
both for our observing program and archival {\em Spitzer} data.

\subsection{Spitzer IRAC Observations}\label{sec:iracobs}

IRAC observations were centered on the NED coordinates of the
sample galaxies based on catalog names listed in
\citet{1993ApJS...89....1R}. A beamsplitter and supporting
optics centers the target on two detectors simultaneously
\citep{2004ApJS..154...10F}, either at 3.6 and 5.8 \m\ or 4.5 and
8.0 \m, and so each observation consisted of two pointings at a common
orientation of the focal plane relative to sky. These observations
were taken as snapshots with no attempt to mosaic or dither. To guard
against saturation, we used the high-dynamic range (HDR) mode which
provides 0.6~s and 12~s integrations at each pointing.

The data were initially processed and calibrated by the IRAC basic
calibration data (BCD) pipeline, version S14.0. The pipeline performs
basic processing tasks, including bias and dark current subtraction;
response linearization for pixels near saturation; flat-fielding based
on observations of high-zodiacal background regions; saturation and
cosmic-ray flagging, the latter indicated by signal detections more
compact than the PSF; and finally flux calibration based on
observations of standard stars\footnote{Details are provided in the
  IRAC Data Handbook, available at
  http://ssc.spitzer.caltech.edu/irac/dh/.}. The nominal photometric
stability for compact sources is better than 3\% in all detectors
\citep{2005PASP..117..978R}. Corrections for extended
sources\footnote{http://ssc.spitzer.caltech.edu/irac/calib} are
accurate to $\sim 10$\%, but the contribution of extended emission and
the resulting correction is small for most of the sample galaxies.
Color-corrections are typically much greater, especially in the 8.0
\m\ channel where in-band PAH emission can result in factors of two or
greater corrections.

\begin{figure}
\epsscale{0.6}
\plotone{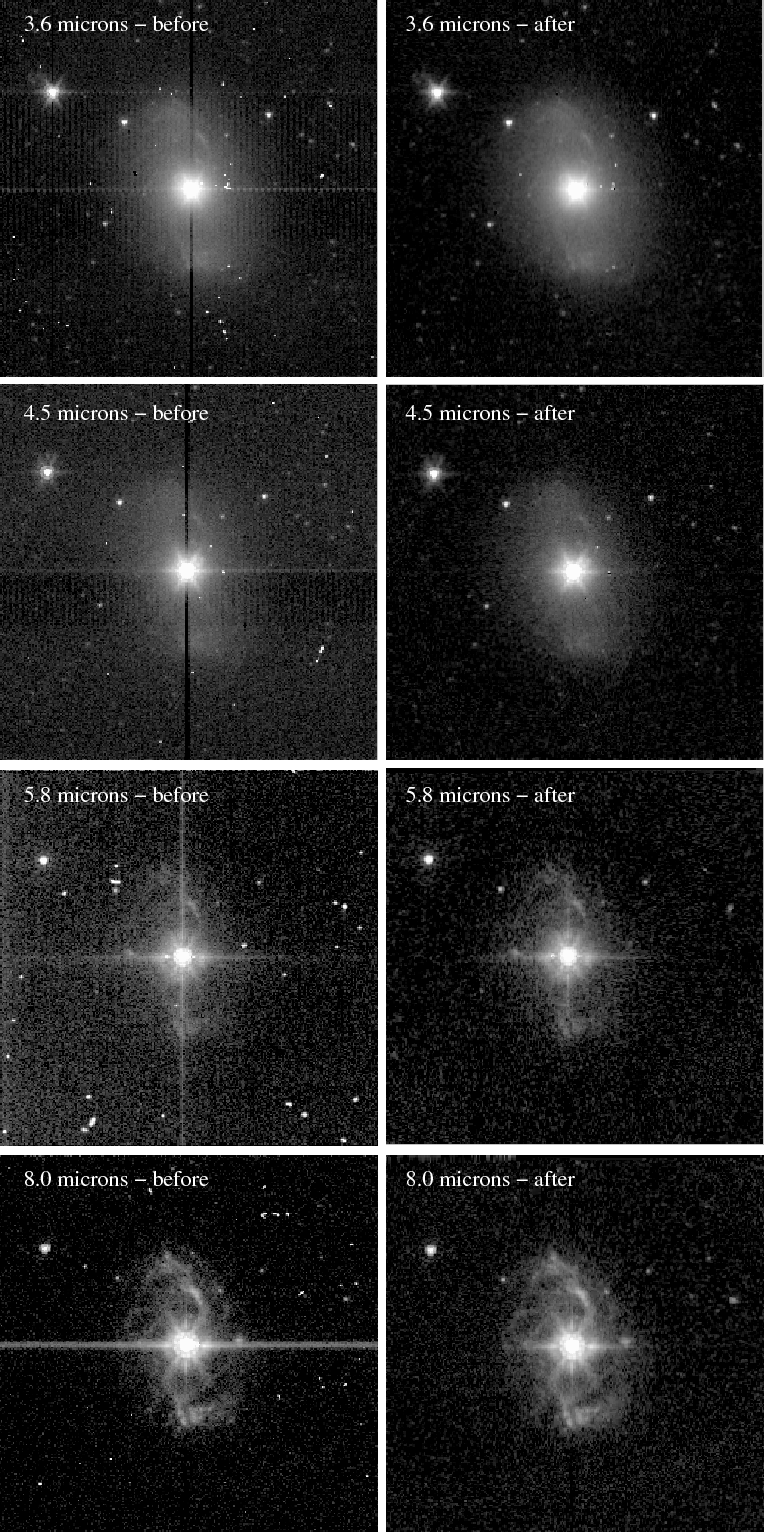}
\caption{IRAC images of NGC~4151, shown to illustrate the results of our post-BCD
  artifact removal technique. Each row corresponds to a single IRAC detector,
  stacked in order of increasing wavelength. The left column shows the
  S14.0 pipeline image, and the right column shows the matching image
  after artifact removal. The images are displayed with a log stretch
  from 0 to 1 MJy sr\mone\ for the 3.6~\m\ and 4.5~\m\ images and 0 to
  20 MJy sr\mone\ for the 5.8 and 8.0~\m\ images, where zero surface
  brightness has been adjusted roughly to the image
  background.  }\label{fig:iracprocess}
\end{figure}

The data were further processed for remaining artifacts, including
cosmic-rays and detector artifacts not corrected by the BCD
pipeline. The steps performed for artifact removal and photometric
extraction are detailed below in
Sections~\ref{sec:cosmicray}--\ref{sec:lastiracstep}. Figure~\ref{fig:iracprocess}
illustrates the effect of our artifact removal techniques.

The IRAC photometry is listed in Table~\ref{tab:iracphot}. Detailed
below, the photometry is presented in the IRAC magnitude system with
zero-point flux densities 280.9, 179.7, 115.0, and 64.13~Jy for the
  3.6, 4.5, 5.8, and 8.0 \m\ channels, respectively
  \citep{2005PASP..117..978R}. The photometry includes corrections for
  extended emission and color-corrections. 

\subsubsection{Cosmic-ray and bandwidth-effect mitigation}\label{sec:cosmicray}

All of the images were affected to varying degrees by cosmic-ray and
solar proton hits. Moreover, 5.8 and 8.0 \m\ images near saturation
suffered from the bandwidth effect, which manifests as a row-wise
trail of fading source images repeating every four pixels from the
affected source. These multiple images do not conserve flux but
artificially add signal to the sources; for our purposes, these
bandwidth effect artifacts behave like cosmic ray hits near our target
galaxies. 

The BCD pipeline attempts to identify cosmic-ray hits by locating
detections that are narrower than the PSF. Extended cosmic ray tracks
are however missed by this procedure. The v. S14.0 BCD pipeline
further has no means of mitigating the bandwidth effect.

We corrected these artifacts in three, post-BCD steps. Firstly, we
generated difference maps between the long and short exposure
images. Difference signal that exceeded 5$\sigma$ on the long exposure
images was flagged as an artifact, and these artifacts were replaced
with signal from the short exposure image. The matching mask and
uncertainty images were updated accordingly. This technique was
particularly successful at removing bandwidth artifacts with little
image degradation; the affected image regions are at relatively high
signal-to-noise on the short exposure images.

We next applied van Dokkum's \citeyearpar{2001PASP..113.1420V} algorithm to
flag cosmic rays not picked up on difference images. Images are
convolved with a Laplacian filter, which enhances sharp edges on image
features and effectively identifies cosmic ray tracks. The IRAC point
response function (PRF) is however undersampled in all four detectors,
and real, compact sources such as field stars or the AGN appear as
false positives. These false positives can be filtered by measuring
the asymmetry of the detected source, parameter $f_{lim}$ in van
Dokkum's notation. Cosmic ray hits tend to be more asymmetric than the
PRF, corresponding to a larger value of $f_{lim}$. Using a few sample
images and trial-and-error, we determined lower threshold values for
$f_{lim}$ that flag obvious cosmic rays but pass field stars and
galaxy images; specifically, we found $f_{lim} = 8$ worked well for
the 4.5, 5.8, and 8.0 \m\ images, and $f_{lim} = 10$ for the 3.6 \m\
images. 

Finally, the few remaining cosmic rays were flagged interactively by
inspection of four-color images. Residual cosmic rays appear as
color-saturated pixels in this representation and so were easily
identified, flagged, and replaced by bilinear interpolation of
neighboring pixels.

\subsubsection{Bias artifacts}

Residual bias artifacts affect the pipeline-processed data. Software
is currently available at the Spitzer Science Center to mitigate these
artifacts where the image comprises mainly compact sources, but the
algorithm breaks down in the presence of extended or diffuse
emission. The host galaxy is detected for many of the sample AGNs, and
so we had to develop new techniques to eliminate these bias artifacts.

The 3.6 and 4.5 \m\ images show the effects of column pull-down and
multiplexer bleed (``muxbleed''). Column pull-down is evident as a
depressed bias level along columns that run through pixels near
saturation. The bias adjustment is nearly constant along a column, but
there may be slightly different bias offsets above and below saturated
pixels.

One approach is to evaluate the bias depression in source-free
regions, but, for some sources, the presence of extended emission over
a majority fraction of the array reduces or eliminates valid
background regions. We suppressed the diffuse emission by performing
row-wise median filtering across pull-down columns and used the median
difference to determine the bias correction.

Muxbleed also affects rows containing pixels near saturation, most
evident as a row pull-up, but also impacting the bias level on
neighboring rows. The images are read as four separate readout
channels that are interlaced every four columns. The bias on each
readout channel is affected differently, resulting in a vertical
pinstripe pattern over some region of the array near saturated pixels.

To mitigate the muxbleed artifact in the presence of extended
emission, we first median-filtered the image using an $8 \times 8$
kernel to smooth out the pinstripe pattern over two readout cycles,
subtracted the median-filtered image to remove diffuse emission, and
generated residual images of each readout channel. Muxbleed appears on
each channel readout image as decaying, horizontal stripes along rows
containing pixels near saturation with surrounding bands of weaker,
constant bias offset. We determined a final muxbleed model by fitting
the brighter muxbleed stripes with a cubic polynomial and determining
the median DC-level offset in the surrounding bands.

The 5.8 \m\ images were affected by residual dark current, appearing
as a slowly varying surface brightness gradient of the
background. This ``first frame'' artifact results from the sensitivity
of the dark current to properties of the previous observation and the
time elapsed since that observation. To remove this dark current
gradient, we first masked bright stars and diffuse emission from
galaxies and then fit a bilinear surface brightness model to the
background. 

Several 5.8 and 8.0 \m\ images were also affected by banding, which is
a decaying signal along rows or columns containing saturated
pixels. These bands were reduced by fitting separate row-wise and
column-wise polynomials to the surface brightness of off-source
regions of the array.

\subsubsection{Saturation and Distortion}

Final corrections were performed using custom IDL scripts and the {\em
  MOPEX} software
package\footnote{http://ssc.spitzer.caltech.edu/postbcd/mopex.html}
\citep{2005ASPC..347...81M, 2006SPIE.6274E..10M}. The short exposure
images were used to replace saturated pixels on the long-exposure
images. The images were then corrected for distortion and registered
to a 1\farcs22 $\times$ 1\farcs22 grid. The resulting data products
are the science image, calibrated in surface brightness units
MJy~sr\mone; an uncertainty image based on the propagation of
statistical uncertainties through the pipeline and post-BCD
processing, but which does not include systematic uncertainties
associated with calibration; and a coverage image, which, for these
snapshot exposures, marks good pixels as ``1,'' bad pixels (i.e.,
known bad pixels or cosmic ray hits) as ``0,'' and intermediate values
indicating that good and bad pixels were used in the distortion
correction for that pixel.

\subsubsection{Photometric extraction}\label{sec:lastiracstep}

We extracted flux density measurements using a synthetic, 20\arcsec\
diameter circular aperture centered on the brightest infrared source
associated with the active galaxy. The exception is NGC~1097, which
has an off-nucleus star-forming region that is brighter than the
central point source; in that case, the aperture was centered on the
central point source. To determine the point-source contribution to
the aperture, each image was convolved with a 2-D, rotationally
symmetric Ricker wavelet \citep{2006MNRAS.369.1603G}. The width of the
central peak of the wavelet was tuned to match the width of the
nominal IRAC PRF, effectively subtracting extended emission and
enhancing point sources. We used field stars to calibrate the point
source response of the wavelet-convolved image; this calibration
includes a correction for aperture losses. We next applied extended
source corrections\footnote{http://ssc.spitzer.caltech.edu/irac/calib/} to
the residual signal (total $-$ point source). 


Finally, the resulting photometry was color-corrected following the
prescription described in the IRAC Data Handbook. The uncorrected
calibration provides a flux density measurement at a nominal
wavelength assuming a nominal spectral shape $\nu F_{\nu} = $~constant
over the broadband response. The color-correction adjusts the
calibration based on the true shape of the spectrum. The result is the
true flux density at a nominal wavelength rather than a broadband
average; for example, the color-corrected flux reported for the
8~\m\ camera will be closer to peak of the 7.7~\m\ PAH feature plus
any underlying continuum at that wavelength rather than
bandpass-weighted average. For reference, the nominal wavelengths for
the IRAC cameras are 3.550, 4.493, 5.731, and 7.872~\m.

Color-correction involves integrating the infrared spectrum
weighted by the IRAC filter bandpasses. To perform this integration,
we used our IRS spectra (Section~\ref{sec:irsobs}, below) with
extrapolation to shorter wavelengths assuming a power law slope
matching the observed, uncorrected 3.6 \m\ flux density. Note that
this color-correction technique does not force a match between the IRS
and IRAC photometry; the flux scale of the spectrum is normalized so
that only the shape of the IRS spectrum influences the color
correction.  Photometric corrections for extended sources are not
accurately known for the IRS spectra, and so the accuracy of the IRS
spectral extractions precluded separating the color corrections for
extended and point source contributions for a given source; rather,
the color correction was determined based on the integrated signal in
the aperture.

\subsection{Archival Spitzer IRAC Observations}\label{sec:iracarchive}

IRAC observations of 16 sample galaxies were obtained through public
release to the {\em Spitzer} archive. Data processing largely followed
the techniques described above for our observations, except that,
where our observations comprise snapshots, the archival observations
all employed dithering or mosaicking techniques. The archival data
were initially mosaicked using standard procedures with {\em MOPEX},
including background matching based on overlaps. 

Artifacts, particularly banding, seriously affected the matching of
mosaic overlap regions and so we developed a modified mosaicking
procedure building on the algorithm of \citet{1995ASPC...77..335R}. We
modified their least-squares approach to employ least absolute
deviations as a more robust measure of the quality of overlap
matching, and we further masked bright source and artifact regions
prior to overlap matching. The final images were assembled using a
median stack of astrometrically aligned images, background subtracted,
and finally sub-imaged to roughly 5\arcmin\ square to match our survey
observations.

Residual bias artifacts were removed by computing column or row biases
where sufficient background was available for bias
determination. Cosmic rays were largely mitigated by the median stack,
but residual bad pixels were identified and flagged interactively, as
described in Section~\ref{sec:cosmicray}.

\subsection{Spitzer IRS Spectral Mapping}\label{sec:irsobs}

We observed the sample galaxies using the {\em Spitzer} IRS in
spectral mapping mode and using the Short-Low (SL) and Long-Low (LL)
modules. The modules cover a wavelength range of $\sim 5$ -- 36 \m\
with resolution $R=\lambda / \Delta \lambda$ ranging from 64 to
128. The integration times were 6 seconds per slit pointing. 

The spectral maps were constructed to span $> 20\arcsec$ in the
cross-slit direction and centered on the target source. The
observations were stepped perpendicular to the slit by half slit-width
spacings. For the SL data, the mapping involved 13 observations
stepped by 1\farcs8 perpendicular to the slit, and, for the LL data, 5
observations stepped by 5\farcs25 perpendicular to the slit. The
resulting spectral cubes span roughly $25\farcs2 \times 54\farcs6
\times$ (5.3 -- 14.2 \m) for the SL data and $29\farcs1 \times
151\arcsec \times$ (14.2 -- 36 \m) for the LL data.

The raw data were processed through the {\em Spitzer} BCD pipeline,
version S15.3.0. The pipeline handled primary processing tasks
including identification of saturated pixels; detection of cosmic ray
hits; correction for ``droop,'' in which charge stored in an
individual pixel is affected by the total flux received by the
detector array; dark current subtraction; and flat-fielding and
response linearization. Details are provided in the IRS Data
Handbook\footnote{http://ssc.spitzer.caltech.edu/mips/dh}. 

Sky subtraction used off-source orders. In SL and LL observations, the
source is centered in the first or second order separately, with the
``off-order'' observing the sky at a position offset parallel to the
slit: 79\arcsec\ away for SL and 192\arcsec\ away for LL. Sky frames
were constructed using median combinations of the off-source data and
subtracted from the on-source data of matching order. No detectable
contamination of the sky frame appeared in our observations based on
inspection of the sky frame data. In addition, a portion of the first
order slit spectrum appears on the second order observation and is
used as an additional check of spectral features that appear near the
first / second order spectral boundary. 

Data cubes were constructed by first registering the individual,
single-slit spectra onto a uniform grid (slit-position
vs. wavelength). The IRS slits do not align with the detector grid,
and the detector pixels undersample the spatial
resolution. Interpolating the undersampled slit onto a uniform grid
produces resampling noise \citep{1989MNRAS.238..603A}, which
significantly impacts the spectra of unresolved sources. We instead
used the pixel re-gridding algorithm described by
\citet{2007PASP..119.1133S} that minimizes resampling noise. Pixels
from the original image are effectively placed atop the new,
registered pixel grid. Uniform surface brightness is assumed across
the original pixel, and that original signal is weighted and
distributed based on the fractional overlap with pixels on the new
grid. After registering the single-slit frames, the data were
re-gridded by bilinear interpolation, one wavelength plane at a time,
into the final data cube. Flux uncertainty images were similarly
processed, with modifications to accommodate variance propagation, to
produce an uncertainty cube. 

The spectra were extracted from synthetic, 20\arcsec\ diameter
circular apertures centered on the brightest, compact IR source nearest
the target coordinates. Fractional pixels at the edge of the aperture
were accounted for by assuming uniform surface brightness across the
pixel and weighting by the area of intersection between the pixel and
aperture mask. 


The cube-extracted spectra were optimally weighted based on a
three-dimensional modification of Horne's
\citeyearpar{1986PASP...98..609H} two-dimensional slit extraction
algorithm. Successful application of this algorithm requires
estimation of spatial profiles $P_{x\lambda}$, the probability that a
detected photon falls in a given pixel $x$ rather than some other
pixel on the spectral map at wavelength $\lambda$. Optimal extraction
requires that the fractional uncertainty of the estimator for $P_{x\lambda}$ is
less than the fractional uncertainty of original spectral image. Generation of
$P_{x\lambda}$ therefore requires some smoothing along the wavelength
axis. To help preserve real variations of $P_{x\lambda}$ with
wavelength, we employed Savitzky-Golay
\citeyearpar{1964AnalyticalChemistry...36..1627} polynomial smoothing. In
this smoothing scheme, $P_{x\lambda}$ is calculated by a polynomial
fit to the spectrum at pixel $x$ over the fitting window $\delta
\lambda$ either centered on $\lambda$ or limited by the ends of the
spectrum. We performed trial-and-error smoothing experiments on
CGCG381-051, which shows relatively weak continuum at short
wavelengths but high eqw PAH features, to decide on the polynomial
order and window smoothing parameters; ultimately we selected
quadratic polynomials and 5-pixel smoothing windows to balance
improved signal-to-noise and the tracking of real variations
of $P_{x\lambda}$ with wavelength. 

\begin{figure}
\epsscale{1.0}
\plotone{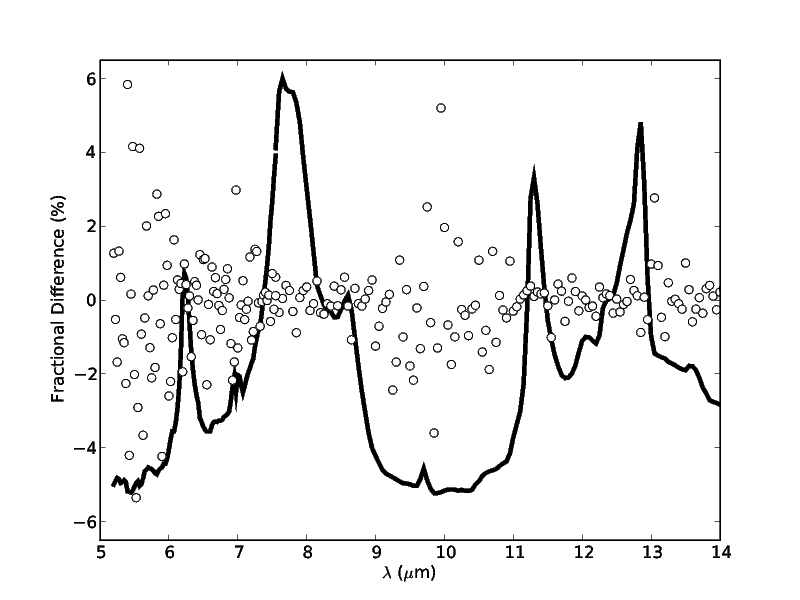}
\caption{An illustration of IRS optimal spectral extraction for
  NGC~3079. The wavelength axis is in the observed frame. The fractional
  differences between the optimally-weighted and non-weighted spectra,
  calculated as a fraction of the optimally-weighted spectrum, are
  plotted as open circles. The optimally-extracted spectrum, $F_{\nu}$, has been
  scaled to fit the plotting range and is shown as a thick line. Even
  though this is an extended source and further shows extended, strong
  PAH features, optimal weighting has preserved the shape and strength
  of the PAH 6.2~\m\ feature. The relative differences typically fall
  in the range 2--4\%, comparable to the statistical uncertainties of the
  non-weighted extraction. In this high signal-to-noise case, optimum
  extraction reduced the statistical uncertainties by $\sim 60\%$. 
}\label{fig:optimumplot1}
\end{figure}

\begin{figure}
\epsscale{1.0}
\plotone{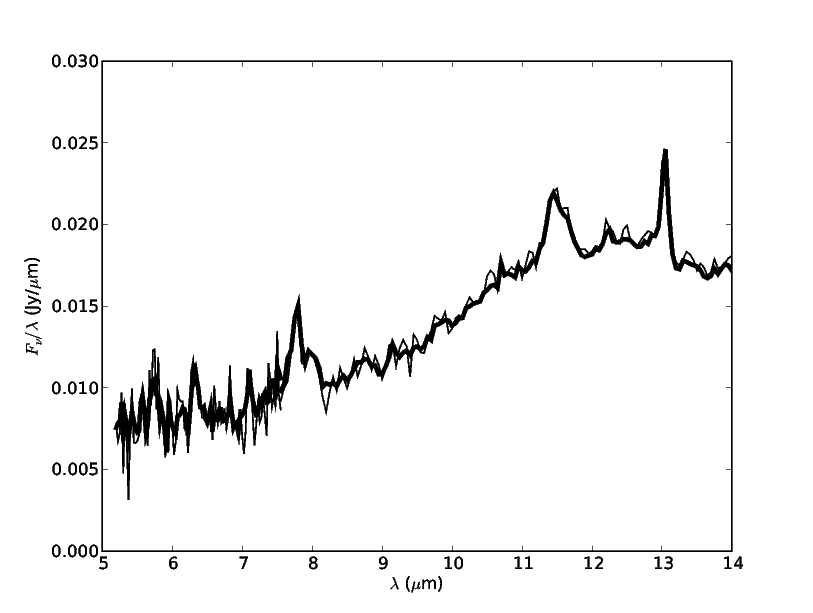}
\caption{An illustration of IRS optimal spectral extraction. The IRS
  short-low spectrum of F01475-0740 is shown. The wavelength axis is
  in the observed frame. The optimally-weighted spectrum is plotted as a
  thick line, and non-weighted spectrum as a thin line. Notice that
  the continuum and PAH features retain their shape with optimal
  weighting; however, the formal statistical uncertainties are reduced by a
  factor of 2--3 over the SL spectral range.
}\label{fig:optimumplot2}
\end{figure}

The effect of optimal extraction is illustrated in
Figures~\ref{fig:optimumplot1} and \ref{fig:optimumplot2}, which
compare the optimally-extracted and non-weighted extraction of the IRS
SL spectrum of NGC~3079 and F01475-0740, and Table~\ref{tab:optnoopt},
which compares the measurements derived from these
extractions (see Section~\ref{sec:pahfit} for a description of the
measurement technique). NGC~3079 presents a challenging case because it is
edge-on and shows bright, extended PAH emission. The spatial profile
is therefore a complex function of wavelength, and coarse smoothing in
the wavelength direction could potentially affect the measurement of
the PAH fluxes and equivalent widths. We find however that the fractional
difference between the optimally-weighted spectrum and the unweighted
spectrum is typically only a few~\%, comparable to the statistical
uncertainties in the unweighted spectrum. Furthermore, the measured
fluxes and equivalent widths agree to within the measurement
uncertainties; in fact, the measurement uncertainties are dominated by
the systematics of the measurement technique.

Compared to NGC~3079, F01475-0740 is a compact source at relatively
low signal-to-noise. Figure~\ref{fig:optimumplot2} clearly illustrates the
advantage of optimum weighting. The continuum shape and PAH spectral
features are preserved, but the formal statistical uncertainties are
reduced by factor of 2--3 over the SL spectral range. Again, the
fluxes and equivalent widths of lines agree to within the measurement
uncertainties, but the optimally weighted spectrum produced
signficant ($>3\sigma$) detections of PAH~6.2~\m, \ion{Ne}{5}~14.3~\m,
and \ion{S}{3}~18.7~\m\ that are too faint for the unweighted
spectrum. 

\begin{figure}
\epsscale{1.0}
\plotone{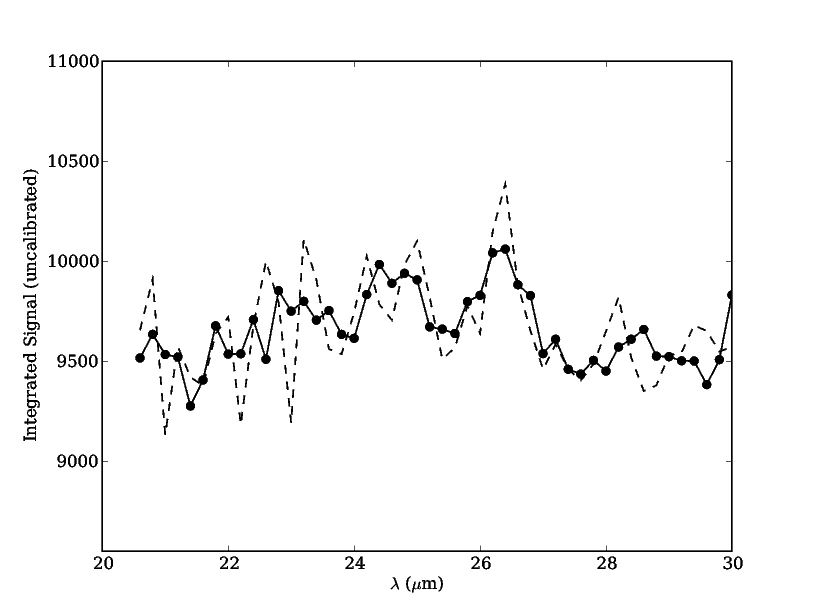}
\caption{An illustration of spectral fringe reduction. Shown is the extracted
  spectrum of MRK~1239, both with (filled circles connected by
  solid-lines) and without (dashed-line) fringe reduction. Wavelengths
  are in the observed frame. The data shown here are not yet
  flux-calibrated, as fringe mitigation is performed prior to flux
  calibration in our processing. Note that the fringes were fit to
  wavelength ranges avoiding spectral lines; the apparent
  [\ion{O}{4}] $\lambda$26~\m\ line in the uncorrected data is
  spurious. }\label{fig:fringeremoval}
\end{figure}

Fringing at the 5\% -- 10\% level was apparent in the LL observations,
particularly those of sample objects with a strong point source
contribution. The spectra undersample the fringing, and so the fringe
pattern could not be removed using a conventional filtering in Fourier
space. We employed instead the technique described by
\citet{2003ESASP.481..375K}, which involves fitting sinusoids in
wavenumber to line-free sections of the spectrum. Successive fringe
components are added to the model until the next model fringe
amplitude falls below the noise level of the spectrum. The effect of
this technique is illustrated in Figure~\ref{fig:fringeremoval}.

The flux scale was calibrated against archival staring and spectral
mapping observations of the stars HR7341 (SL) and HR6606 (LL). The
staring mode spectra were extracted using
SPICE\footnote{http://ssc.spitzer.caltech.edu/postbcd/spice.html.},
and the spectral mapping observations were processed into cubes as
described above for our sample galaxies. The flux calibration curve
was determined by the ratio of the flux-calibrated staring mode
spectra to the uncalibrated, cube-extracted spectra and smoothed by a
polynomial fit. 

No correction was produced for resolved, extended emission. The flux
calibration includes an aperture loss correction factor appropriate
for compact sources, but extended sources do not suffer as much
aperture loss. Therefore, wavelength bands containing extended
emission relative to the PRF will be calibrated systematically
high. Discussed below in Section~\ref{sec:iracirs}, the systematic
error introduced is of order 10\%, particularly near the 8 \m\ PAH
features. 

We note that these IRS observations have been independently and
differently processed and recently presented in
\citet{2009ApJ...701..658W}. The main differences in data processing
are: (1) Wu et al. used the CUBISM tool and its calibrations for
spectral cube construction \citep{2007PASP..119.1133S}; (2) it is not
clear that Wu et al. extracted the spectra optimally; (3) a
rectangular aperture, which covers a similar solid angle to our
extraction aperture, was commonly used (20\farcs4 $\times$ 15\farcs3;
larger for SINGS galaxies); (4) there appears to have been no attempt
to de-fringe the long wavelength (LL) data; (5) for an unknown number
of sources, Wu et al. had to scale the SL data to match up with the
flux calibration of the LL data. We address these differences in turn,
excepting (3), the minor difference of extraction aperture geometry.

We found that (1) CUBISM forced the data onto a new grid in which the
original pointing center was shifted, commonly resulting in nuclei
that were centered between pixels rather than on a pixel. This shift
apparently introduced a systematic error in the resampling of the
surface brightness, because we noticed artifacts and a few percent
reduced signal in the spectrum extracted from cubes produced by
CUBISM. Our cubing technique was designed to require minimal
resampling of the observational grid and mitigated these artifacts.

We also found that (2) non-optimal extractions decreased the
signal-to-noise of the extracted spectra of fainter sources by factors
of $\sim 50\%$ or more (see Figure~\ref{fig:optimumplot2} and the
discussion above). Although it varies source to source, (4) fringing
can produce $\sim 10\%$ artifacts between 14 and 36~\m\ (cf. their
extraction of MRK~1239 and M-6-30-15, among others). Finally, (5) we
did not need to apply any additional scaling to the spectra extracted
from SL cubes; rather our calibrations of the SL module show excellent
agreement with both the IRAC measurements and LL measurements, even
though the calibrations for each respective {\em Spitzer} module were
based on different calibration stars; this result suggests that the
calibration against spectrally mapped stars, as presented here, was
more robust for calibration and systematic corrections, accepting the
$\sim 10\%$ calibration uncertainty for sources dominated by extended
emission.

\subsection{Archival IRS Data}\label{sec:irsarchive}

Archival data of our sample galaxies, such as obtained by the SINGS
collaboration, were processed in the same manner as for our program
data. Spectral mapping data were available in LL for all sources, but
there are however a few sources for which only SL staring mode
observations are available: NGC~526A, NGC~4941, NGC~3227, IC~5063,
NGC~7172, and NGC~7314. The staring mode spectra were extracted using
SPICE and subjectively scaled to achieve the best match to align with
LL measurements and, where possible, IRAC 5.8 and 8.0
\m\ measurements. Not surprisingly, there result disagreements between
the IRAC measurements and scaled, IRS staring observations owing to
the contribution of the host galaxy in the 20\arcsec\ aperture.

\subsection{Spitzer MIPS-SED Observations}\label{sec:mipsobs}

The sample galaxies were observed using the 70 \m\ low-resolution
spectrometer (SED mode) of the MIPS instrument. The spectrometer
provides spectral resolution $R\sim 15$--25 between $\lambda$55 and 95
\m. The slit dimensions are $20\arcsec \times 120\arcsec$.

Each observation consisted of three pairs of on-source and off-source
measurements, where the off-position was located 1\arcmin\ --
3\arcmin\ from the on-source position. Integration times were 3
or 10 seconds depending on the IRAS 60 \m\ flux density of the
source. The observations include measurements of built-in calibration
light sources, called stimulators, for flux calibration.

We used post-BCD products of the S14.4.0 pipeline; the pipeline
processing includes flux calibration, background subtraction,
co-addition of the on-source pointings, and uncertainty
images. Details are provided in the MIPS data
handbook \footnote{http://ssc.spitzer.caltech.edu/mips/dh}.

The MIPS spectra were extracted by summing across a 29\arcsec-wide
synthetic aperture centered on the target source (three-column
extraction). The spectra were further corrected for aperture losses
assuming a point source model; the present extractions include no
corrections for spatially extended emission. The photometric accuracy
is expected to be 10\% for compact sources and $\sim 15$\% for
extended sources \citep{2008PASP..120..328L}. 


Note that the MIPS aperture covers a solid angle $\sim 1.8\times$
larger than the IRAC and IRS apertures. Modeling SEDs of sources with
extended ($> 20\arcsec$) far-infrared emission will therefore require
an (unknown) aperture correction that is potentially much greater than
the reported calibration uncertainty. Candidates for MIPS aperture
corrections will show a jump in the MIPS flux relative to a suitable
extrapolation of the IRS spectrum, although a real spectral peak in
the 40--50~\m\ range might similarly result in an observed MIPS SED that
appears too blue even in the absence of aperture effects.

\subsection{Comparison of IRS and IRAC Photometry}\label{sec:iracirs}

\begin{figure}
\plottwo{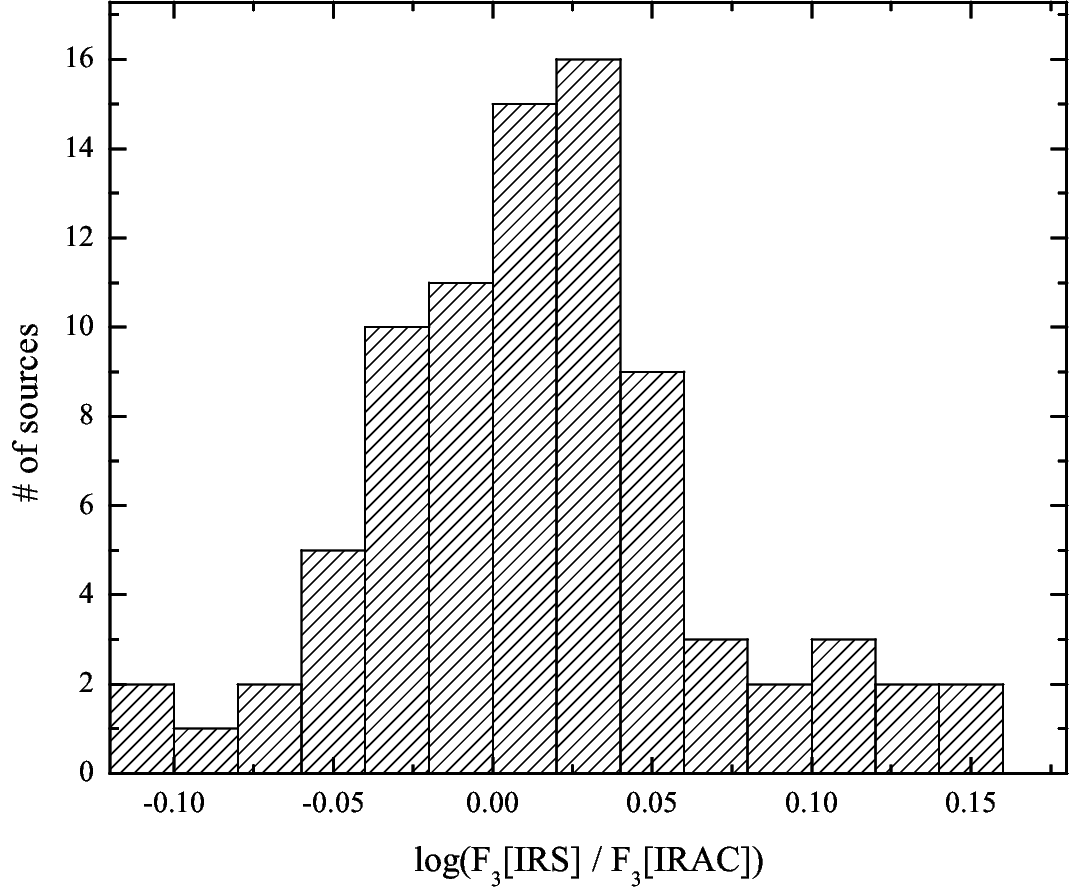}{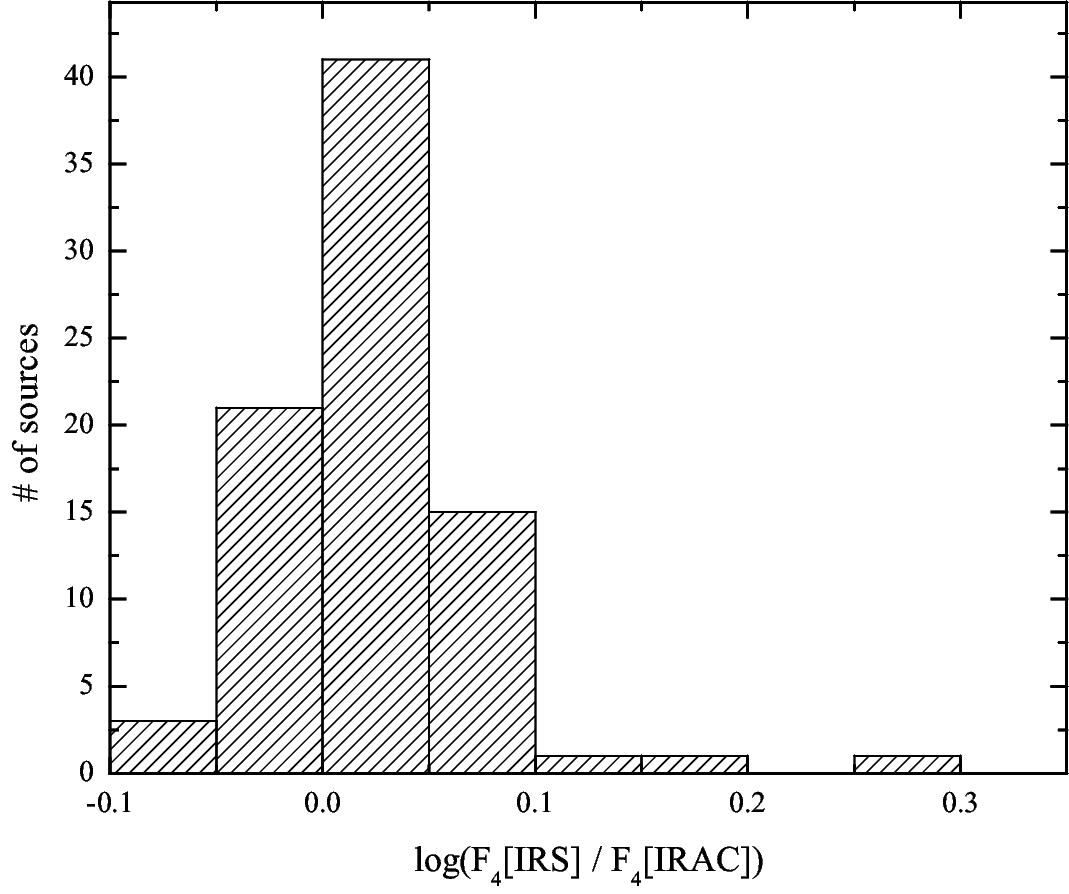}
\caption{Frequency distribution of the flux density ratio
  $F_{\nu}({\rm IRS}) / F_{\nu}({\rm IRAC})$ at the overlap
  wavelengths 5.8 \m\ (IRAC channel 3 = 5.731 \m) and 8.0 \m\ (IRAC
  channel 4 = 7.872 \m).}\label{fig:irsiracoverlap}
\end{figure}

The IRS spectra overlap with the 5.8 and 8.0 \m\ IRAC channels. To
compare the relative spectrophotometry, we interpolated the IRS
spectra to find the flux densities at the effective wavelengths of the
color-corrected IRAC data, 5.731 \m\ and 7.872 \m, respectively. The
average relative difference of the overlapping flux densities,
$F_{\nu}({\rm IRS}) / F_{\nu}({\rm IRAC}) - 1$, are $4\pm 1$\% at 5.8 \m\ and
$6\pm 1$\% at 8.0 \m. The frequency distributions are shown in
Figure~\ref{fig:irsiracoverlap}. For comparison, the average standard
scores, $(F_{\nu}[{\rm IRS}] - F_{\nu}[{\rm IRAC}])/\sigma$, are
$0.00\pm 0.03$ and $0.18 \pm 0.05$. 

These results indicate that the IRS flux densities are, on average,
systematically higher than the IRAC measurements by a few
percent. However, the excess at 5.8 \m\ is dominated by statistical
uncertainties, as the average standard score is consistent with zero.

Strong, extended PAH emission however contributes significantly to the
8.0 \m\ IRAC channel. This spectral feature is difficult to
color-correct accurately owing to numerical integration errors that
arise at sharp spectral features. In addition, PAH emission is often
spatially resolved in this sample, and the present IRS calibration
includes no correction for extended emission. The systematic offset of
6\% may result from the overcorrection of aperture losses for extended PAH
sources. The IRAC photometry of extended sources is moreover accurate
to only $\sim 10$\%. 

\begin{figure}
\plotone{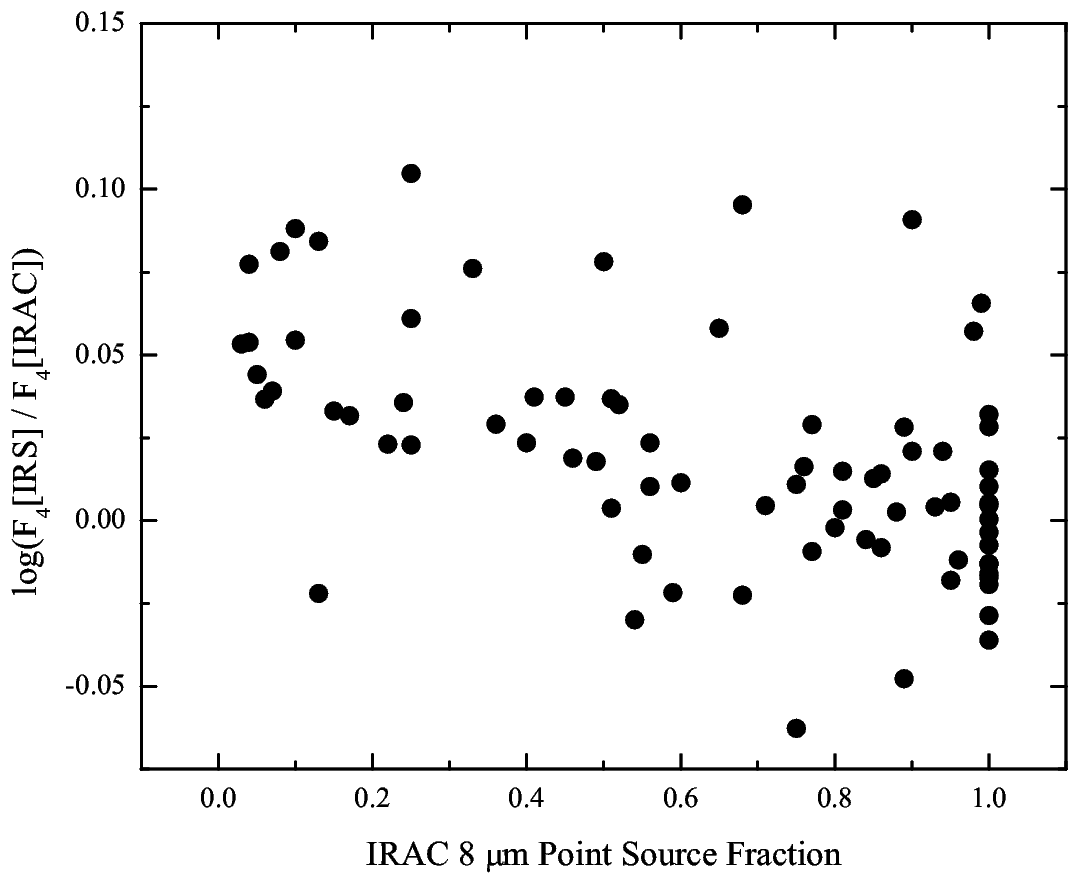}
\caption{IRS / IRAC 8 \m\ flux ratio vs. IRAC point source fraction. The
  point source fraction is the ratio of the point source flux
  determined by wavelet convolution to the total flux in the
  20\arcsec\ synthetic aperture. Here, the subscript ``4'' refers to
  IRAC channel 4 = 8 \m.}\label{fig:eightmicronexcess}
\end{figure}

Figure~\ref{fig:eightmicronexcess} shows the 8 \m\ IRS / IRAC flux ratio as
a function of the fractional contribution of the central point source
to the 20\arcsec\ nuclear aperture. The fainter sources provide
appreciable scatter, but the brighter sources reveal a trend in which
point-source dominated objects show better agreement, but more
extended objects present $\sim 10$\% IRS excesses. This trend supports
the interpretation that residual extended source calibration
uncertainties are the primary cause for the IRS -- IRAC discrepancies
at 8 \m. 

\section{Analysis}\label{sec:analysis}

\subsection{Spectrum Decomposition}\label{sec:pahfit}

\begin{figure}
\plottwo{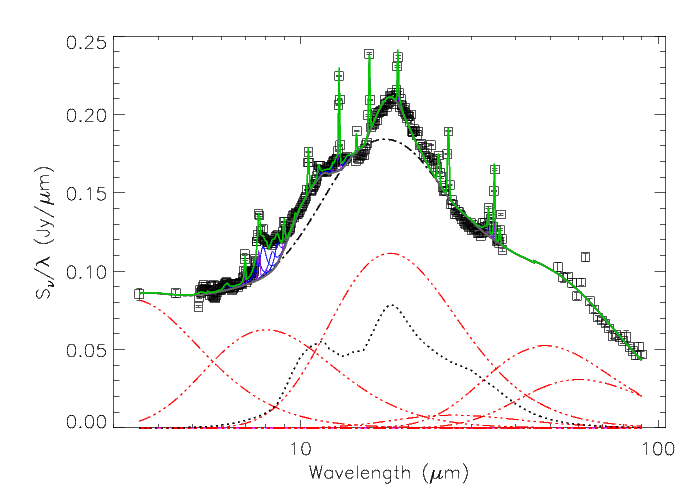}{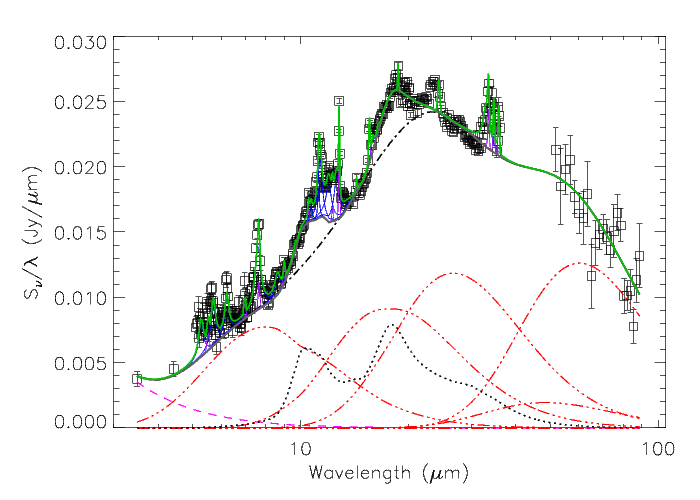}
\caption{Example spectrum decomposition for NGC~4151 (left) and
  NGC~7213 (right). The vertical
  axis is in $\nu F_{\nu}$ units. The measurements are represented by
  square symbols; the uncertainties are plotted as vertical error-bars
and are usually smaller than the symbol size. Continuous lines
represent the best fit model (green), underlying dust continuum
(gray), and individual PAH features (blue) and emission lines
(purple). The broken lines mark the following components: stars (pink
dashed), dust continuum (red dash-dot-dot-dot), and optically-thin, warm dust emission
(black dotted). The baseline used for Sil index measurements (stars +
dust continuum + warm, thin dust continuum) is shown as the black
dash-dot-dash line.}\label{fig:n4151decomposition}
\end{figure}

In addition to continuum radiation from dust grains and associated Sil
emission and absorption bands, the infrared spectrum of Seyfert
galaxies includes diagnostics such as fine structure lines tracing a
range of ionization states, \molhyd\ lines, and PAH emission. We used
the spectrum fitting tool PAHFIT \citep{2007ApJ...656..770S}, which is
tailored to low resolution IRS spectroscopy. The fits to the SEDs of
NGC~4151 and NGC~7213 are provided in
Figure~\ref{fig:n4151decomposition} as examples of the
decomposition. The results are summarized in the following tables:
integrated PAH fluxes are given in Table~\ref{tab:pahstr}; PAH
equivalent widths (EQWs) in Table~\ref{tab:paheqw}; H$_2$ line fluxes
(mostly upper limits) in Table~\ref{tab:h2str} and EQWs in
Table~\ref{tab:h2eqw}; ionic fine structure line fluxes in
Table~\ref{tab:fsstr}, and their EQWs in Table~\ref{tab:fseqw}. The
line measurements are not corrected for model extinction, but for
completeness we list the best-fit model dust opacity (normalized to
10~\m) in Table~\ref{tab:tauapcor}.

As provided, PAHFIT is best suited for nearly normal or star-forming
galaxies; its model includes thermal continuum radiation from dust
grains, PAH features, fine-structure lines from lower ionization state
species, \molhyd, and Sil absorption, whether by assumed mixed or
foreground screen extinction. We added two components to the PAHFIT
model to fit the Seyfert SEDs: (1) fine-structure lines from high
ionization states, such as [\ion{Ne}{5}] and [\ion{Ne}{6}]; and (2),
to fit silicate emission features, a simple model for warm dust clouds
that are optically thin at infrared wavelengths.

By default, PAHFIT models extinction based on the dust opacity law of
\citet{2004ApJ...609..826K}. \citet{2008ApJ...678..729S} found that
the cold dust model of \citet{1992A&A...261..567O} better matches the
high 18 \m\ Sil / 10 \m\ Sil absorption found in active ultraluminous
infrared galaxies. From inspection, the present spectra similarly show
relatively strong 18 \m\ features, whether in emission or absorption,
and so we further modifed PAHFIT to use the cold dust model of
\citet{1992A&A...261..567O}. 

The thin, warm dust model assumes clouds with simple, slab geometry
and opacity at 10 \m, $\tau_{10} < 1$. These warm clouds are further
assumed to be partially covered by cold, absorbing dust clouds. The
model spectrum is then, 
\begin{equation}
F_{\nu} = \left(1 - C_f\right) B_{\nu}(T_W)\left(1 - e^{-\tau_W(\nu)}\right) + C_f
B_{\nu}(T_W) \left(1-e^{-\tau_W(\nu)}\right)  e^{-\tau_C(\nu)} , \label{eqn:hotdustmodel}
\end{equation}
where $C_f$ is the covering fraction of cold, foreground clouds, modeled
independent of the galaxy extinction; $\tau_W$ is the opacity through
the warm clouds; $\tau_C$ is the opacity through the cold, foreground
clouds, and $B_{\nu}(T_{w})$ is the source function, for which we
adopt a scaled Planck spectrum at (fitted) temperature $T_W$ for
simplicity. Note that the cold dust opacities described by
Eq.~\ref{eqn:hotdustmodel} are taken to be independent of the global
PAHFIT dust opacity model; the opacity values listed in
Table~\ref{tab:tauapcor} are determined by the global dust opacity
fit. The intention of including this additional model component
is to provide a realistic continuum baseline for emission line fits
and Sil strength determination rather than an interpretable, radiative
transfer model, and details of the opacity law or more realistic
source functions are beyond the scope of the present spectral
decomposition.


We further modified PAHFIT better to accommodate the IRAC broadband
and MIPS-SED measurements. Both instruments provide much lower
sampling density in wavelength than IRS data, and consequently they
receive lower weight in the $\chi^2$ minimization procedure. We
therefore employed the sampling weight correction described by
\citet{2007ApJ...670..129M}, which effectively re-weights individual
data points based on the sampling density local to that data point;
regions of low sampling density receive increased weight. The
weighting is normalized so that each data point carries, on average,
unity sampling weight. We also introduced as a fitted parameter an
aperture correction factor for the MIPS-SED wavelength range that
boosts the model SED by a factor up to 1.81, which is the areal
aperture ratio of the MIPS-SED extraction to the nominal
20\arcsec\ diameter extraction aperture.  Table~\ref{tab:tauapcor}
includes the best-fit aperture corrections.

There are a few caveats to the results of this decomposition. These
low spectral resolution data are not best suited for measuring line
fluxes, particularly in spectrally crowded regions. PAHFIT takes a
conservative approach, severely restricting the centroid wavelengths
and widths of the fitted lines; for example, the fine-structure lines
are assumed to be unresolved and their widths are fixed at the
instrumental resolution. Even so, the fine-structure lines near 35
\m\ are crowded, and furthermore the IRS measurements are
very noisy at that region of the spectrum, with sensitivity $\sim
10\times$ poorer than at 25 \m, depending on background
(\url[http://ssc.spitzer.caltech.edu/documents/som/]{Spitzer
  Observer's Manual}). These limitations are reflected in the
uncertainties and upper limits reported in
Tables~\ref{tab:pahstr}--\ref{tab:fseqw}. 



Extracted line fluxes were however affected by occasional, residually
poor fits to the local continuum or PAH features. For example, if the
PAHFIT model placed the continuum too low locally to an emission line,
PAHFIT would grow the emission line to meet the data and therefore
produce a line flux greater than observed. Similarly, the PAHFIT model
might produce a local continuum that is too high and the line flux is
reported too low. A good illustration of this problem is the too-high
continuum model surrounding the [\ion{Ne}{5}] $\lambda$14.3~\m\ line
of NGC~4151 (Figure~\ref{fig:n4151decomposition}). To compensate for
locally poor continuum models, the mean and rms of the
model-subtracted spectrum was evaluated in spectral regions
surrounding each line. Each fitted line peak was accordingly adjusted
by subtracting the local, mean residual continuum. Line fluxes were
similarly adjusted; line widths are unaffected as they were held fixed
to the instrumental resolution during the fit.


There is a weak PAH feature near 14.3~\m\ that potentially
contaminates the [\ion{Ne}{5}] $\lambda$14.3~\m\ fine structure line
\citep{2000A&A...358..481S}. Deblending these features uses the fact
that the PAH feature is somewhat broader (FWHM $\sim 0.4$~\m) than the
unresolved (FWHM $< 0.1$~\m) [\ion{Ne}{5}] line
\citep{2007ApJ...656..770S}. In the worst-case scenario, the code
may fail to fit a weak but present PAH 14.3 feature resulting in an
artificially enhanced [\ion{Ne}{5}] line strength. This effect is at
least partially ameliorated by the residuals analysis and is reflected
in the large uncertainties of Table~\ref{tab:fsstr}. The [\ion{Ne}{5}]
14 / 24 ratio provides however a good check for contamination.  This
ratio is a density diagnostic with lower limit $\sim 0.9$
\citep[e.g.,][]{1999ApJ...512..204A}. We demonstrate in a companion paper (Baum
et al. 2009, submitted) that for all of the sources where there is a
[\ion{Ne}{5}] 14.3~\m\ detection, the line ratio is consistent with
the low density limit \citep[cf.][]{2002A&A...393..821S}; contamination from
PAH 14.3 would push the ratio to a (forbidden) value below the low
density limit. We conclude that the PAH~14.3~\m\ feature does not
significantly contaminate the [\ion{Ne}{5}] $\lambda$14.3~\m\ line
strengths in this study or that any contamination is within
the uncertainties of the line strength.

\subsubsection{Silicate Strengths}\label{sec:silicates}

This spectrum decomposition technique provides a reasonable model for
the 9 -- 20 \m\ continuum, suitable to
measure the relative strength of Sil features.
\citet{2007ApJ...654L..49S} defined the Sil strength as the log ratio
of the observed flux density at the center of the Sil feature, 10 \m\
or 18 \m, and the local continuum; e.g.,
\begin{equation}
S_{10} = \ln\left(\frac{F_{10 \mu{\rm m}}[{\rm
      observed}]}{F_{10 \mu{\rm m}}[{\rm continuum}]}\right).
\end{equation}

To measure the Sil strength for the 12 \m\ sample, we first subtracted
PAH and other emission line features as determined by PAHFIT to obtain
$S_{10 \mu{\rm m}}[{\rm observed}]$. The continuum was derived from
the spectrum decomposition as the sum of the (optically thick) dust
components, stars, and the continuous part of the warm, thin dust
component; the Sil emission features of the warm, thin dust component
were replaced by quadratic interpolation between bracketing spectral
regions. We used Monte Carlo variation of the PAHFIT model parameters
and the data uncertainties to determine the Sil strength
uncertainties.

The measured Sil strengths are provided in Table~\ref{tab:silicates}.

\subsubsection{Continuum Spectral Indices}\label{sec:indices}

We further used the PAHFIT spectral decomposition to produce line-free
continuum spectra over $\lambda$20--30 \m. The MIPS SED data are
similarly line-free, except for a few possible detections of
\ion{O}{1} ($\lambda$63 \m); see, e.g.,
Figure~\ref{fig:n4151decomposition}. 

The data show a range of continuum slopes, and we characterized the
spectral shape by fitting a powerlaw model, $F_{\lambda} \propto
\lambda^{\alpha}$ where $\alpha$ is the spectral index, to the rest
wavelength ranges 20--30 \m\ and 55--90 \m. The results are listed in
Table~\ref{tab:spectralindices}. Note that in this convention for
$\alpha$, the Rayleigh-Jeans tail of the Planck spectrum would give
$\alpha = -4$. 


\subsubsection{Comparison with Measurements Employing Spline
  Approximations for the IR Continuum}

\citet{2009ApJ...701..658W} adopt a different but conventional
approach to the measurement of the PAH 6.2~\m\ and 11.2~\m\ features and
the 10~\m\ Sil strength, and we next consider systematic differences
with our measurements. Rather than decompose the spectrum with a
dust and lines model as PAHFIT does, their approach was to define a local
continuum level based on a spline fit to wavelength ranges narrowly
bracketing PAH features. To measure Sil strengths, they adopted the
technique of \citet{2007ApJ...654L..49S}, which requires the
identification of apparently feature-free continuum points to anchor a
broader spline interpolation across the Sil features.

\citet{2007ApJ...656..770S} demonstrated that, for the nearly normal
galaxies in the SINGS sample, the PAH line strengths measured by
PAHFIT are systematically greater, by factors of 2--3, than line
strengths that are based on a spline fit to the neighboring
pseudo-continuum. The reason is qualitatively illustrated in the
PAHFIT decomposition of NGC~7213
(Figure~\ref{fig:n4151decomposition}). Both the 6.2~\m\ and
11.3~\m\ features blend with weaker, overlapping PAH
features. By defining the continuum level based on neighboring
spectral points without accounting for PAH blending, the continuum
level is overestimated, and the line strength and eqw are
underestimated. It is further evident from this decomposition that Sil
strengths will be systematically affected if the influence of PAH
blends and the underlying continuum shape are not reasonably accounted
for; it would not be surprising that the spline technique of produced
very different Sil strengths particularly for this source.

\begin{figure}
\plotone{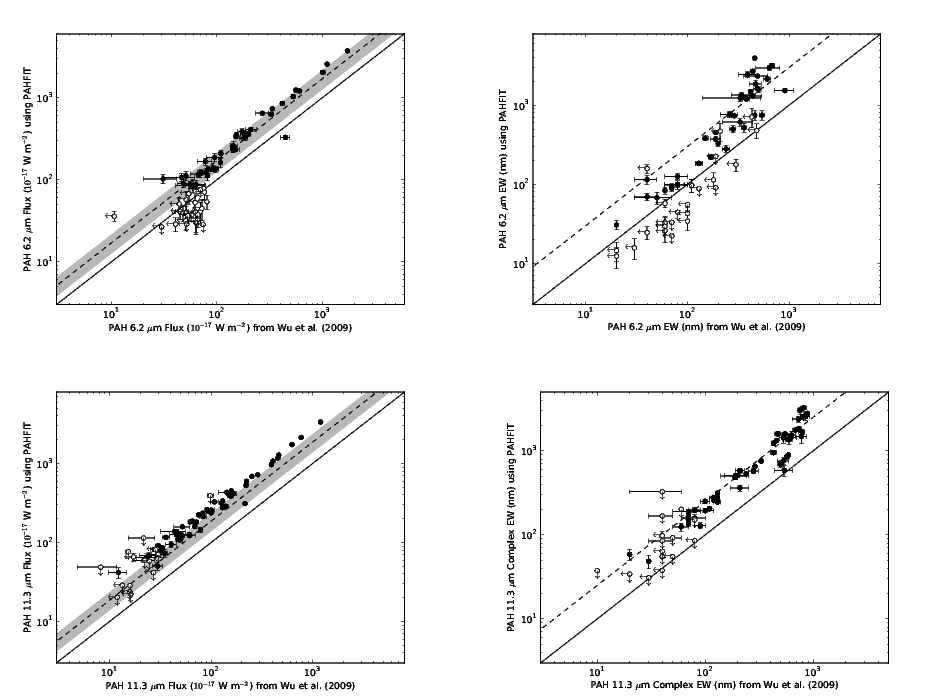}
\caption{Comparison of the PAHFIT spectrum decomposition analysis
  ($y$-axis on these plots) and the local continuum analysis of
  \citet{2009ApJ...701..658W} ($x$-axis on these plots). Only data
  with nearly matching apertures are included in these plots.
  The
  individual comparisons are {\em top left}: PAH 6.2~\m\ flux, {\em
    top right}: PAH 6.2~\m\ eqw, {\em bottom left}: flux of the PAH
  11.3~\m\, which includes the 11.22~\m\ and 11.33~\m\ PAH features, 
  and {\em bottom right}: PAH 11.3~\m\ complex eqw. Data that include upper
  limits are indicated by open circles, and detections are marked by
  filled circles. The solid line indicates loci of equivalent
  measurements. On the flux-flux diagrams, shaded regions bisected by
  a dashed line mark the average ratio and $\pm 1\,\sigma$ range
  reported by \citet{2007ApJ...656..770S} for galaxies in the SINGS sample.  On the eqw
  diagrams, the dashed line illustrates a factor of 3 (6.2~\m) or  2.5
  (11.3~\m) enhancement for PAHFIT
  compared to the spline technique; this line is purely for
  illustration and is not based on a fit to the data. Notice that the
  PAHFIT-derived fluxes and eqws are, on average, systematically
  higher, because PAHFIT removes the contamination of neighboring,
  weak PAH features.}
\label{fig:wupahcomparison}
\end{figure}

\begin{figure}
\plotone{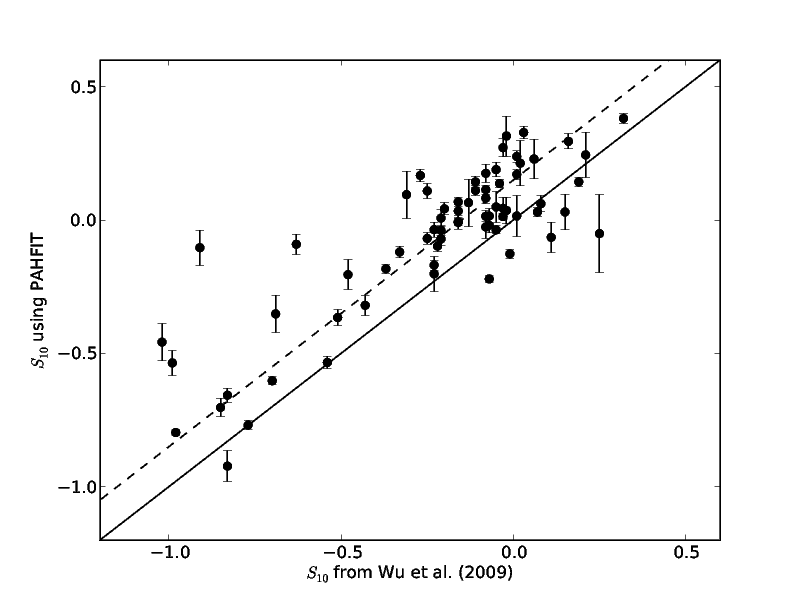}
\caption{Comparison of Sil~10~\m\ feature strengths based on the
  PAHFIT spectrum decomposition analysis 
  ($y$-axis) and the spline continuum approximation ($y$-axis) used by 
  \citet{2009ApJ...701..658W}. The solid line indicates loci of equivalent
  measurements, and the dashed line illustrates a 0.15 dex enhancement of
  Sil strengths measured by PAHFIT compared to the spline technique;
  this line is purely for illustration and is not based on a fit to
  the data.}\label{fig:wusilcomparison}
\end{figure}

We illustrate the systematic differences between the present analysis
and that of \citet{2009ApJ...701..658W} in
Figs.~\ref{fig:wupahcomparison} and \ref{fig:wusilcomparison}. As
expected, PAHFIT measures, on average, systematically higher values of
PAH fluxes and eqws, because PAHFIT removes the contamination of
neighboring PAH features to measurement of the local continuum. The
measurements of \citet{2009ApJ...701..658W} fall somewhat below the
average ratio (PAHFIT / spline continuum) reported by
\citet{2007ApJ...656..770S}, but the spline measurements will be
sensitive to systematic differences in how the local continuum anchor
points were defined in the analysis; such a detailed reconciliation is
beyond the scope of the present work.

Similarly, our PAHFIT-derived Sil strengths are systematically more
positive, by $\sim 0.1$--0.3 dex, indicating weaker Sil absorption or
stronger Sil emission depending on the sign of the Sil
strength. Again, this result is unsurprising, because PAH blends that
are not accounted for by decomposition can falsely mimic enhanced
continuum surrounding the Sil~10~\m\ feature, pushing the Sil strength
to lower (more negative) values. Recall that we also use an
interpolation technique similar to that of
\citet{2007ApJ...654L..49S} to measure the strength of
the Sil features; the difference is that we perform the analysis on
the line-subtracted continuum model produced by PAHFIT.

\subsection{SEDs Averaged by Optical Classification}

\begin{figure}
\plotone{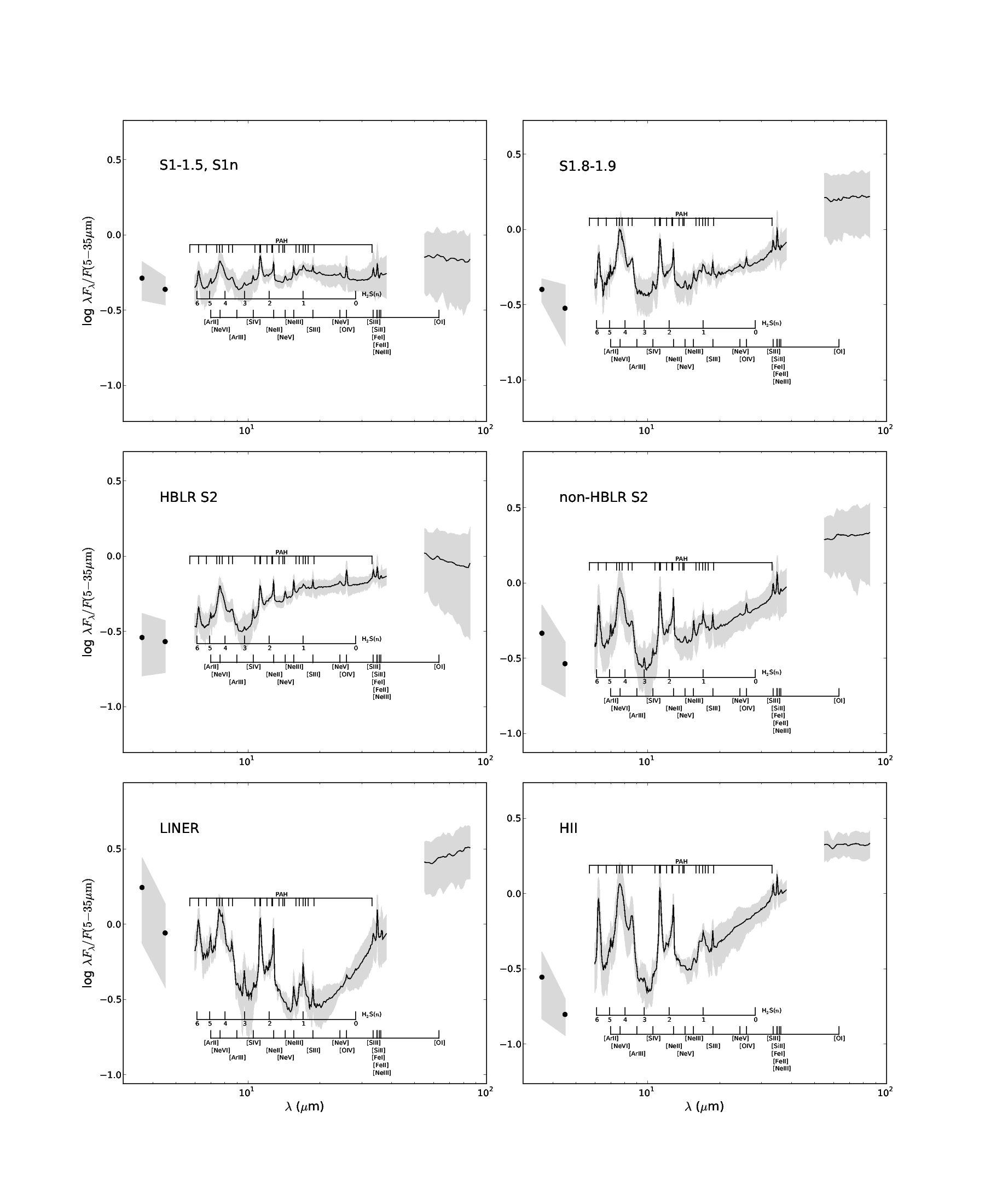}
\caption{SEDs
  averaged by optical classification. All SEDs were 
  normalized to $F$(5--35\m) prior to averaging.  The
  gray filled regions indicate the median absolute deviation among
  objects within that classification bin. PAH, \molhyd, and
  fine-structure lines included in the PAHFIT spectral decomposition
  are annotated. }\label{fig:avgdSEDs}
\end{figure}

A key goal of this project is to identify and compare the infrared
characteristics of AGNs segregated by optical classification. Toward a
qualitative first look, we calculated average SEDs within the
following classification bins: (1) S1.0-1.5 \& S1n; (2) S1.8-1.9; (3)
HBLR Seyfert 2s (S1h \& S1i); (4) non-HBLR
Seyfert 2s (S2); (5) LINERs; and (6) HII. The results are presented in
Figure~\ref{fig:avgdSEDs}. The separation of the non-HBLR and HBLR S2s
was motivated by inspection of Figure~\ref{fig:allseds}; non-HBLR S2s
appear to have higher-eqw PAH features, for example. Recall that the
non-HBLR S2s may in fact harbor a BLR that has not appeared in
spectropolarimetric or infrared measurements. All but three of the 20
non-HBLR S2s have been searched for an HBLR, but the
inclusion of these three sources in the non-HBLR subsample 
does not appear to dilute the striking differences between the averaged
spectra of non-HBLR S2s and HBLR S2s.

To perform the averaging, all of the data were corrected for redshift
and interpolated to a common wavelength grid. The SEDs were converted
to $\lambda F_{\lambda}$ and normalized to the flux density integrated
between 5 and 35~\m, $F$(5--35\m). Objects within a given
classification bin were averaged, and the median absolute deviation
was computed as a robust estimator of the characteristic spread of
SEDs within a classification bin.

\citet{2003ApJ...583..632T} demonstrated that S1s and HBLR S2s show
similar IRAS broadband colors, but non-HBLR S2s tend to show cooler
IRAS colors \citep[cf.][]{1997Natur.385..700H}. The present study
confirms this result in some finer detail based on the averaged
SEDs. From inspection of Figure~\ref{fig:avgdSEDs}, S1s (group~1) and
HBLR S2s (group~3) show the flattest infrared SEDs. They both present
fine-structure emission lines of high-ionization state species, such
as [\ion{O}{4}], [\ion{Ne}{4}], and [\ion{S}{4}], with similar equivalent
width. Hidden S1s have a slightly redder SED, on average, and also
show evidence for Sil 10\m\ absorption.

Similarly, the average SEDs of S1.8-1.9s (group 2) and non-HBLR S2s
(group 4) are essentially indistinguishable. Both groups show strong
PAH features, red continuum, and blue IRAC colors suggesting
significant contribution of stellar photospheric emission at the short
wavelength end. The SEDs of these groups most closely resemble
optically classified star-forming galaxies, or starbursts (HII; group
6), except that the average HII SED for this sample shows PAH features
with somewhat greater equivalent widths. The [\ion{S}{3}] and
[\ion{Si}{2}] average equivalent widths appear comparable between
S1.8-1.9, non-HBLR S2, and HII galaxies.

The average LINER SED stands out by showing a bowl-shaped infrared
SED. The SEDs appear to be more strongly dominated by stellar
photospheres, or perhaps very hot dust, at shorter wavelengths
compared to the other groups, including HII galaxies where starlight
appears to dominate the IRAC bands. In this way, the LINERs in our
survey are similar to the IR-faint LINERs in the larger sample studied by
\citet{2006ApJ...653L..13S}. This result is somewhat tempered
by the broad range of IRAC colors observed among the LINERs in this
survey; from inspection of the six individual LINER SEDs, four show
bowl-like SEDs resembling the average (NGC~2639, NGC~4579, NGC~4594,
\& NGC~5005), and two show HII-like SEDs (NGC~1097 \& NGC~3079).

\begin{figure}
\plotone{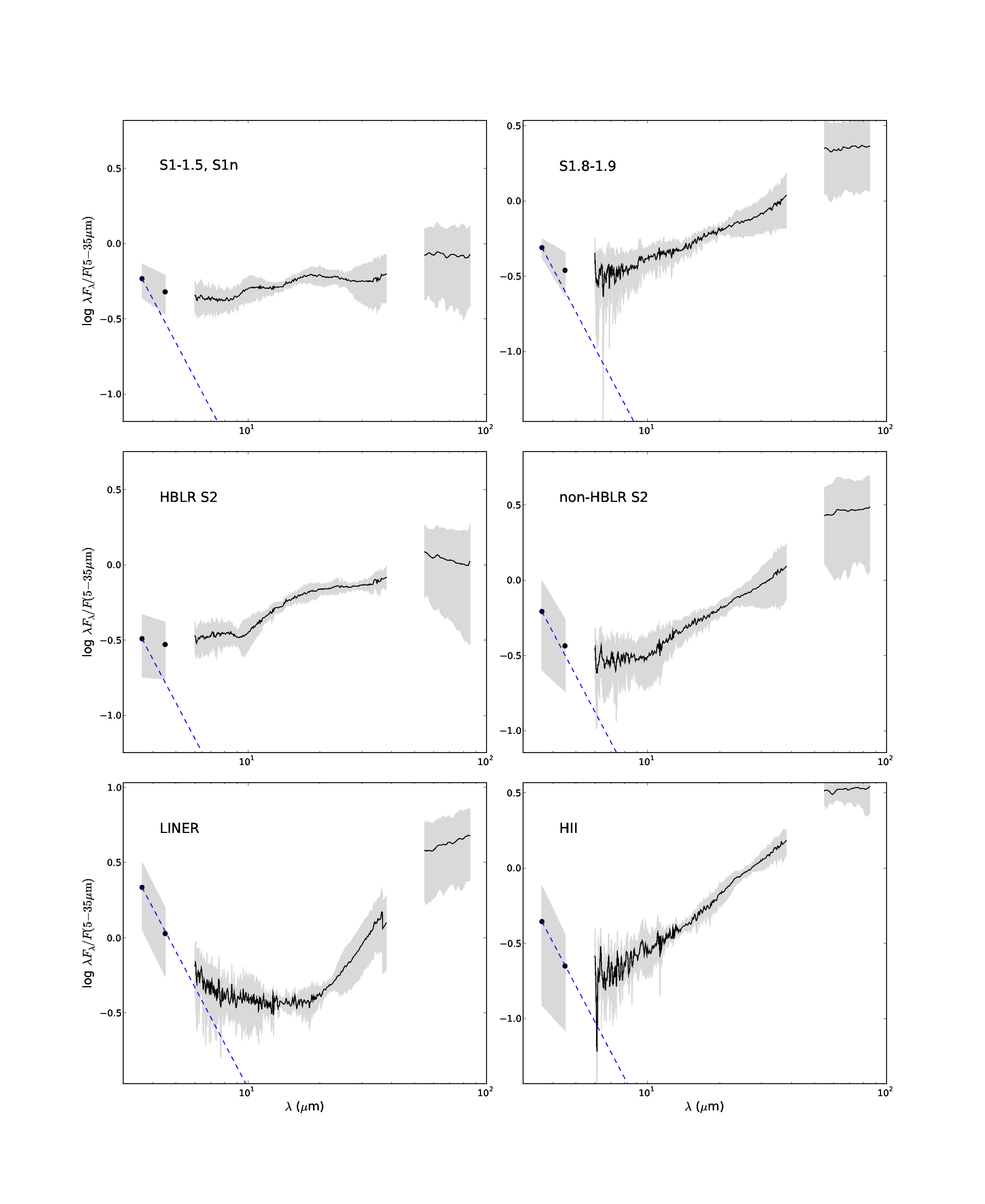}
\caption{PAH, \molhyd, and fine-structure line-subtracted SEDs
  averaged by optical classification. All SEDs were 
  normalized to $F$(5--35\m) prior to averaging.  The
  gray filled regions indicate the median absolute deviation among
  objects within that classification bin. The dashed lines trace
  a Rayleigh-Jeans continuum spectrum, representing an approximate
  spectrum for stellar photospheres, anchored to 3.6 \m. Sil
  features appear in emission for the S1-1.5 \& S1n class but in
  absorption in the averaged SEDs both the HBLR S2s (S1h and
  S1i) and of non-HBLR S2s (S2s where a HBLR has not yet been
  detected). }\label{fig:avgdLineSubs}   
\end{figure}

PAH features are commonly detected in this sample, and, even though
such features appear at reduced equivalent width in S1 and HBLR S2
objects \citep[cf.][]{2000A&A...357..839C}, they are sufficiently
strong to obscure Sil features, especially Sil emission. We therefore
repeated the SED averaging after subtracting PAH, \molhyd, and
fine-structure lines based on the results of the PAHFIT decomposition
(Section~\ref{sec:pahfit}); the results are provided in
Figure~\ref{fig:avgdLineSubs}. Here the average SED of S1 and HBLR S2
distinguish more clearly, with S1 showing clear 10 \m\ and 18 \m\ Sil
emission features, similar to that observed in QSOs
\citep{2005ApJ...625L..75H}. In contrast, the averaged SED of known
HBLR S2s show Sil 10 \m\ in absorption.

The other classes again show broadly similar SEDs after line
subtraction. The notable exception is the non-HBLR S2 average, where
weak 10 \m\ Sil absorption appears underneath the subtraction of very
strong PAH features. Sil features are essentially absent among
intermediate Seyferts (S1.8-1.9), LINERs, and HII galaxies. It is
further interesting to note that the IRAC color [3.6] $-$ [4.5] of the
averaged S2, LINER, and HII SEDs is consistent with an undiluted
Rayleigh-Jeans continuum. The averaged SEDs of the other classes
present redder [3.6] $-$ [4.5], indicating dilution from warm dust or
some flat-spectrum component.

\section{Discussion}\label{sec:conclusions}

We have presented the data reduction and decomposition of {\em
  Spitzer} Space Telescope 3.6 -- 90~\m\ spectrophotometry of active
galaxies from the extended 12~\m\ survey. Careful attention was
provided to matching 20\arcsec, circular diameter apertures across
the IRAC and IRS bands (3.6 -- 36~\m) with appropriate color and
extended source corrections where possible or with an evaluation of
the systematic error where such corrections were not available. 

We further present SEDs averaged within groups defined by optical AGN
classification. We demonstrate that, within this sample, Seyfert 1s
show Sil emission on average, known HBLR Seyfert 2s show Sil
absorption. This result is broadly compatible with the obscuring torus
interpretation, in which case Seyfert 1s are viewed more nearly
pole-on, affording a more direct view of hot, Sil emitting
dust. HBLR S2s are viewed more nearly edge-on, preferentially
through colder, Sil absorbing dust. That the Sil features are, on
average, weak is further compatible with the clumpy torus model
\citep{2008ApJ...685..160N,
  2008ApJ...678..729S,2008ApJ...685..147N,2007ApJ...654L..45L}.   

The other classes, Seyfert 1.8-1.9, non-HBLR S2, LINER, and HII
galaxies, produce very weak or absent Sil features. They further show
stronger PAH features, bluer IRAC colors, and stronger far-infrared
emission (relative to $F$[5--35\m]). Such SEDs appear to be more
commonly dominated by stellar photospheres and star-forming
processes. Based on the present analysis, however, we are unable to
conclude whether Seyfert 1.8-1.9 and non-HBLR S2 galaxies are in fact
more commonly dominated by star formation or whether this result is
peculiar to the 12~\m\ sample owing to selection effects; for example,
they may harbor less luminous AGNs, or more heavily absorbed AGNs, but
the contribution from star-formation enhanced the 12~\m\ flux density
sufficiently to be included in the 12~\m\ sample. On the other hand,
our results are consistent with the interpretation that the host
galaxy dominates the emission of non-HBLR S2s, diminishing our ability
to detect the HBLR \citep{2001MNRAS.320L..15A}. 

In companion work, we present a statistical analysis of the present
measurements with attention to differences and similarities between
sources grouped by optical classification (Baum et al. 2009,
submitted). We are also investigating a decomposition of the SEDs
using gridded radiative transfer models with the goal of measuring bolometric
contributions of the AGN vs. star-formation as well as constraints on
clumpy torus parameters (Gallimore et al. in prep.). 

\acknowledgments The authors gratefully acknowledge the anonymous
referee for a careful reading of the manuscript and very helpful
comments. This work is based on observations made with
the Spitzer Space Telescope, which is operated by the Jet Propulsion
Laboratory, California Institute of Technology under a contract with
NASA. Support for this work at Bucknell University, the University of
Rochester and, the Rochester Institute of Technology was provided by
NASA through an award issued by JPL/Caltech. A. Yzaguirre received
support from the National Science Foundation REU Program, grant
0097424. J. Jakoboski received support as a Bucknell Presidential
Fellow. 

{\it Facility:} \facility{Spitzer}


\clearpage

\begin{deluxetable}{lllrrrrl}
\tabletypesize{\footnotesize}
\tablewidth{0pt}
\tablecaption{AGNs from the extended 12~\m\  sample observed by the Spitzer Space Telescope\label{tab:sample}}
\tablehead{\colhead{Source ID} & \colhead{Class} & \colhead{Ref} & \colhead{RA (J2000)} & \colhead{Dec (J2000)} & \colhead{cz} & \colhead{D} & \colhead{PID} \\ 
\colhead{} & \colhead{} & \colhead{} & \colhead{(h:m:s)} & \colhead{(\degr:\arcmin:\arcsec)} & \colhead{(km s\mone)} & (Mpc) & \colhead{} } 
\startdata
MRK335      &S1n       & 1 & 00:06:19.53 &    20:12:10.5 & 7730 & 110.4 & 3269 \\
MRK938      &HII       & 2 & 00:11:06.56 & $-$12:06:27.3 & 5881 & 84.0  & 3269, 3672 \\
E12-G21     &S1n       & 3 & 00:40:45.93 & $-$79:14:24.2 & 9000 & 128.6 & 3269 \\
MRK348      &S1h       & 4 & 00:48:47.16 &    31:57:25.2 & 4507 & 64.4  & 3269 \\
NGC424      &S1h       & 5 & 01:11:27.66 & $-$38:05:00.0 & 3527 & 50.4  & 3269 \\
NGC526A     &S1.9      & 6 & 01:23:54.39 & $-$35:03:55.4 & 5725 & 81.8  & 86, 3269 \\
NGC513      &S1h       & 7 & 01:24:26.78 &    33:47:58.4 & 5859 & 83.7  & 3269 \\
F01475-0740 &S1h       & 8 & 01:50:02.69 & $-$07:25:48.4 & 5296 & 75.7  & 3269 \\
\enddata
\end{deluxetable}

\begin{deluxetable}{lrrrrrrrr}
\tabletypesize{\footnotesize}
\tablewidth{0pt}
\tablecaption{IRAC Photometry of the 12 \m\ AGN sample\label{tab:iracphot}}
\tablehead{\colhead{Object} & \colhead{[3.6]} & \colhead{unc} & \colhead{[4.5]} & \colhead{unc} & \colhead{[5.8]} & \colhead{unc} & \colhead{[8.0]} & \colhead{unc}}
\startdata
MRK335  & 8.99 & 0.07 & 8.31 & 0.07 & 7.67 & 0.06 & 6.69 & 0.05 \\
MRK938  & 9.55 & 0.05 & 9.24 & 0.04 & 7.74 & 0.05 & 5.09 & 0.02 \\
E12-G21  & 9.63 & 0.09 & 9.14 & 0.10 & 8.31 & 0.08 & 6.22 & 0.07 \\
MRK348  & 9.27 & 0.08 & 8.46 & 0.08 & 7.67 & 0.06 & 6.38 & 0.04 \\
NGC424  & 7.83 & 0.05 & 6.94 & 0.04 & 6.11 & 0.03 & 4.95 & 0.02 \\
NGC526A  & 9.00 & 0.07 & 8.33 & 0.07 & 7.66 & 0.07 & 6.67 & 0.06 \\
NGC5135  & 8.95 & 0.10 & 8.60 & 0.09 & 7.35 & 0.08 & 4.82 & 0.07 \\
F01475-0740  & 10.83 & 0.16 & 9.89 & 0.14 & 8.59 & 0.10 & 7.16 & 0.06 \\
\enddata
\tablecomments{IRAC photometry is presented in the {\em Spitzer} IRAC magnitude
  system. Zero point flux densities are 280.9, 179.7, 115.0, and 64.13 Jy for the
  3.6, 4.5, 5.8, and 8.0 \m\ channels, respectively. Color corrections
  and corrections for extended emission have been applied. Uncertainties
  include contributions from detector noise and photon counting
  statistics, noise in the IRS spectra and numerical integration errors
  as they contribute to the color correction, and the (10\% of extended
  flux) uncertainty of the extended emission correction. }
\end{deluxetable}

\clearpage

\begin{deluxetable}{lrrrrrrrrr}
\tabletypesize{\scriptsize}
\tablewidth{0pt}
\tablecaption{Comparison of Line Measurements using Optimal and
  Non-optimal Weighting\label{tab:optnoopt}}
\tablehead{
\colhead{Line} & \colhead{$\lambda$(\m)} & \multicolumn{2}{c}{Non-opt. Flux}  &
\multicolumn{2}{c}{Opt. Flux}  & \multicolumn{2}{c}{Non-opt. EQW}  &
\multicolumn{2}{c}{Opt. EQW}\\
\colhead{(1)} & \colhead{(2)} & \multicolumn{2}{c}{(3)}  &
\multicolumn{2}{c}{(4)}  & \multicolumn{2}{c}{(5)}  &
\multicolumn{2}{c}{(6)}\\
}
\startdata
\multicolumn{10}{c}{NGC 3079}\\
H$_2$S(1) & 17.05 & 51 & (5) & 51 & (4) & 137 & (14) & 135 & (10)\\
$[$NeII$]$ & 12.81 & 183 & (11) & 190 & (12) & 345 & (22) & 352 & (22)\\
$[$NeIII$]$ & 15.56 & 22 & (6) & 23 & (6) & 41 & (12) & 44 & (11)\\
$[$OIV$]$ & 25.91 & 30 & (7) & 31 & (6) & 38 & (8) & 39 & (8)\\
$[$SIII$]$ & 18.71 & 18 & (3) & 16 & (3) & 51 & (7) & 44 & (7)\\
$[$SIII$]$ & 33.50 & 63 & (14) & 61 & (15) & 43 & (10) & 42 & (11)\\
$[$SiII$]$ & 34.82 & 260 & (27) & 262 & (26) & 165 & (17) & 166 & (16)\\
\enddata
\tablecomments{Integrated fluxes are given in units of
  $10^{-17}$~W~m\mtwo, and equivalent widths are given in units of
  nm. Uncertainties ($1\sigma$) are listed in parentheses. Upper
  limits are $3\sigma$. Only lines with a detection in the optimal
  extraction are presented; all lines that are detected in the
  unweighted spectrum are also detected in the optimally extracted
  spectrum. Measurements from the unweighted spectrum are given in
  columns (3) and (5); measurements from the optimally weighted
  spectrum are given in the neighboring columns (4) and (6).} 
\end{deluxetable}

\setlength{\tabcolsep}{2pt} 
\begin{deluxetable}{lrrrrrrrrrrrrrrrrrrrrrr}
\rotate
\tabletypesize{\scriptsize}
\tablewidth{0pt}
\tablecaption{PAH Fluxes\label{tab:pahstr}}
\tablehead{\colhead{Source ID} & \multicolumn{2}{c}{6.2 \m}& \multicolumn{2}{c}{7.4 \m}& \multicolumn{2}{c}{7.6 \m}& \multicolumn{2}{c}{7.8 \m}& \multicolumn{2}{c}{8.3 \m}& \multicolumn{2}{c}{8.6 \m}& \multicolumn{2}{c}{11.2 \m}& \multicolumn{2}{c}{11.3 \m}& \multicolumn{2}{c}{12.0 \m}& \multicolumn{2}{c}{12.6 \m}& \multicolumn{2}{c}{17.0 \m}}
\startdata
MRK335          & $< 29 $ & \nodata & $< 105 $ & \nodata & $< 29 $ & \nodata & $ 58 $ & (10) & $ 37 $ & (11) & $ 28 $ & (8) & $< 7 $ & \nodata & $ 22 $ & (6) & $ 27 $ & (6) & $ 31 $ & (4) & $< 18 $ & \nodata \\
MRK938          & $ 735 $ & (34) & $ 1470 $ & (150) & $ 789 $ & (45) & $ 943 $ & (40) & $ 415 $ & (29) & $ 451 $ & (21) & $ 152 $ & (4) & $ 378 $ & (12) & $ 223 $ & (15) & $ 487 $ & (20) & $ 225 $ & (26) \\
E12-G21         & $ 208 $ & (21) & $< 201 $ & \nodata & $ 258 $ & (18) & $ 263 $ & (17) & $ 57 $ & (11) & $ 148 $ & (9) & $ 33 $ & (4) & $ 190 $ & (10) & $ 67 $ & (8) & $ 142 $ & (15) & $ 96 $ & (8) \\
MRK348          & $< 37 $ & \nodata & $ 244 $ & (31) & $< 44 $ & \nodata & $ 113 $ & (16) & $< 39 $ & \nodata & $ 58 $ & (9) & $ 14 $ & (3) & $ 51 $ & (6) & $< 28 $ & \nodata & $ 24 $ & (7) & $< 46 $ & \nodata \\
NGC424          & $ 41 $ & (12) & $ 509 $ & (76) & $ 137 $ & (33) & $ 189 $ & (38) & $ 145 $ & (24) & $ 88 $ & (19) & $< 21 $ & \nodata & $ 55 $ & (14) & $ 57 $ & (14) & $< 56 $ & \nodata & $< 60 $ & \nodata \\
NGC526A         & $ 22 $ & (7) & $ 207 $ & (29) & $ 35 $ & (8) & $ 37 $ & (11) & $ 43 $ & (9) & $ 26 $ & (6) & $< 7 $ & \nodata & $ 30 $ & (6) & $ 45 $ & (5) & $ 26 $ & (5) & $< 31 $ & \nodata \\
NGC513          & $ 186 $ & (20) & $< 312 $ & \nodata & $ 355 $ & (31) & $ 236 $ & (31) & $< 84 $ & \nodata & $ 153 $ & (19) & $ 99 $ & (3) & $ 150 $ & (10) & $ 76 $ & (12) & $ 105 $ & (12) & $ 109 $ & (11) \\
F01475-0740     & $ 40 $ & (9) & $ 90 $ & (29) & $ 103 $ & (10) & $< 33 $ & \nodata & $< 23 $ & \nodata & $ 30 $ & (6) & $ 13 $ & (2) & $ 69 $ & (5) & $ 60 $ & (6) & $ 66 $ & (6) & $< 27 $ & \nodata \\
\enddata
\tablecomments{Integrated fluxes are given in units of $10^{-17}$~W~m\mtwo. Uncertainties ($1\sigma$) are listed in parentheses. Upper limits are $3\sigma$. }
\end{deluxetable}
\setlength{\tabcolsep}{4pt}

\setlength{\tabcolsep}{2pt} 
\begin{deluxetable}{lrrrrrrrrrrrrrrrrrrrrrr}
\rotate
\tabletypesize{\scriptsize}
\tablewidth{0pt}
\tablecaption{PAH Feature EQWs\label{tab:paheqw}}
\tablehead{\colhead{Source ID} & \multicolumn{2}{c}{6.2 \m}& \multicolumn{2}{c}{7.4 \m}& \multicolumn{2}{c}{7.6 \m}& \multicolumn{2}{c}{7.8 \m}& \multicolumn{2}{c}{8.3 \m}& \multicolumn{2}{c}{8.6 \m}& \multicolumn{2}{c}{11.2 \m}& \multicolumn{2}{c}{11.3 \m}& \multicolumn{2}{c}{12.0 \m}& \multicolumn{2}{c}{12.6 \m}& \multicolumn{2}{c}{17.0 \m}}
\startdata
MRK335          & $< 32 $ & \nodata & $< 167 $ & \nodata & $< 48 $ & \nodata & $ 102 $ & (18) & $ 71 $ & (21) & $ 56 $ & (16) & $< 14 $ & \nodata & $ 50 $ & (13) & $ 71 $ & (15) & $ 91 $ & (11) & $< 65 $ & \nodata \\
MRK938          & $ 2700 $ & (130) & $ 5880 $ & (590) & $ 3150 $ & (180) & $ 3730 $ & (160) & $ 1620 $ & (110) & $ 1797 $ & (82) & $ 596 $ & (17) & $ 1429 $ & (44) & $ 622 $ & (41) & $ 1133 $ & (48) & $ 388 $ & (45) \\
E12-G21         & $ 500 $ & (51) & $< 554 $ & \nodata & $ 723 $ & (50) & $ 757 $ & (50) & $ 174 $ & (35) & $ 465 $ & (27) & $ 139 $ & (15) & $ 803 $ & (42) & $ 303 $ & (36) & $ 682 $ & (74) & $ 640 $ & (54) \\
MRK348          & $< 44 $ & \nodata & $ 362 $ & (46) & $< 67 $ & \nodata & $ 181 $ & (25) & $< 66 $ & \nodata & $ 104 $ & (16) & $ 25 $ & (5) & $ 96 $ & (12) & $< 52 $ & \nodata & $ 44 $ & (13) & $< 85 $ & \nodata \\
NGC424          & $ 12 $ & (3) & $ 191 $ & (29) & $ 53 $ & (13) & $ 76 $ & (15) & $ 63 $ & (10) & $ 38 $ & (8) & $< 10 $ & \nodata & $ 26 $ & (6) & $ 30 $ & (7) & $< 31 $ & \nodata & $< 41 $ & \nodata \\
NGC526A         & $ 33 $ & (9) & $ 381 $ & (53) & $ 66 $ & (14) & $ 72 $ & (22) & $ 87 $ & (19) & $ 51 $ & (13) & $< 13 $ & \nodata & $ 57 $ & (11) & $ 96 $ & (11) & $ 58 $ & (11) & $< 87 $ & \nodata \\
NGC513          & $ 613 $ & (67) & $< 1145 $ & \nodata & $ 1330 $ & (120) & $ 900 $ & (120) & $< 337 $ & \nodata & $ 632 $ & (79) & $ 541 $ & (18) & $ 832 $ & (55) & $ 445 $ & (72) & $ 648 $ & (71) & $ 804 $ & (80) \\
F01475-0740     & $ 113 $ & (26) & $ 251 $ & (81) & $ 287 $ & (27) & $< 92 $ & \nodata & $< 65 $ & \nodata & $ 84 $ & (16) & $ 30 $ & (5) & $ 161 $ & (11) & $ 153 $ & (15) & $ 173 $ & (16) & $< 62 $ & \nodata \\
\enddata
\tablecomments{PAH equivalent widths are given in units of nm. Uncertainties ($1\sigma$) are listed in parentheses. Upper limits are $3\sigma$. }
\end{deluxetable}
\setlength{\tabcolsep}{4pt}

\setlength{\tabcolsep}{4pt} 
\begin{deluxetable}{lrrrrrrrrrrrrrrrr}
\tabletypesize{\scriptsize}
\tablewidth{0pt}
\tablecaption{H$_2$ Line Fluxes\label{tab:h2str}}
\tablehead{\colhead{Source ID} & \multicolumn{2}{c}{S(7)}& \multicolumn{2}{c}{S(6)}& \multicolumn{2}{c}{S(5)}& \multicolumn{2}{c}{S(4)}& \multicolumn{2}{c}{S(3)}& \multicolumn{2}{c}{S(2)}& \multicolumn{2}{c}{S(1)}& \multicolumn{2}{c}{S(0)}}
\startdata
MRK335          & $< 72 $ & \nodata & $< 88 $ & \nodata & $< 64 $ & \nodata & $< 55 $ & \nodata & $ 53 $ & (12) & $ 21 $ & (6) & $ 17 $ & (5) & $< 18 $ & \nodata \\
MRK938          & $ 160 $ & (32) & $< 205 $ & \nodata & $ 151 $ & (22) & $ 249 $ & (57) & $ 128 $ & (13) & $ 143 $ & (22) & $ 195 $ & (22) & $< 153 $ & \nodata \\
E12-G21         & $ 55 $ & (15) & $< 132 $ & \nodata & $< 67 $ & \nodata & $< 83 $ & \nodata & $ 40 $ & (11) & $< 34 $ & \nodata & $< 21 $ & \nodata & $< 22 $ & \nodata \\
MRK348          & $< 79 $ & \nodata & $< 85 $ & \nodata & $ 62 $ & (17) & $< 74 $ & \nodata & $< 26 $ & \nodata & $< 28 $ & \nodata & $< 39 $ & \nodata & $ 23 $ & (6) \\
NGC424          & $< 96 $ & \nodata & $< 76 $ & \nodata & $< 68 $ & \nodata & $< 177 $ & \nodata & $< 55 $ & \nodata & $< 54 $ & \nodata & $< 35 $ & \nodata & $< 56 $ & \nodata \\
NGC526A         & $ 76 $ & (19) & $ 40 $ & (11) & $ 73 $ & (10) & $ 55 $ & (11) & $ 29 $ & (9) & $ 22 $ & (4) & $ 33 $ & (9) & $< 28 $ & \nodata \\
NGC513          & $< 94 $ & \nodata & $< 135 $ & \nodata & $< 75 $ & \nodata & $< 137 $ & \nodata & $ 81 $ & (18) & $< 47 $ & \nodata & $ 41 $ & (9) & $< 26 $ & \nodata \\
F01475-0740     & $< 111 $ & \nodata & $< 54 $ & \nodata & $ 47 $ & (11) & $< 49 $ & \nodata & $< 24 $ & \nodata & $ 29 $ & (8) & $< 23 $ & \nodata & $< 43 $ & \nodata \\
\enddata
\tablecomments{Integrated fluxes are given in units of $10^{-18}$~W~m\mtwo. Uncertainties ($1\sigma$) are listed in parentheses. Upper limits are $3\sigma$. The line wavelengths for the S(7) -- S(0) series are 5.512, 6.159, 6.859, 8.004, 9.656, 12.329, 17.063, and 28.171 \m, respectively. }
\end{deluxetable}
\setlength{\tabcolsep}{4pt} 

\setlength{\tabcolsep}{4pt} 
\begin{deluxetable}{lrrrrrrrrrrrrrrrr}
\tabletypesize{\scriptsize}
\tablewidth{0pt}
\tablecaption{H$_2$ EQWs\label{tab:h2eqw}}
\tablehead{\colhead{Source ID} & \multicolumn{2}{c}{S(7)}& \multicolumn{2}{c}{S(6)}& \multicolumn{2}{c}{S(5)}& \multicolumn{2}{c}{S(4)}& \multicolumn{2}{c}{S(3)}& \multicolumn{2}{c}{S(2)}& \multicolumn{2}{c}{S(1)}& \multicolumn{2}{c}{S(0)}}
\startdata
MRK335          & $< 7 $ & \nodata & $< 10 $ & \nodata & $< 9 $ & \nodata & $< 10 $ & \nodata & $ 9 $ & (2) & $ 6 $ & (2) & $ 6 $ & (2) & $< 15 $ & \nodata \\
MRK938          & $ 50 $ & (10) & $< 74 $ & \nodata & $ 60 $ & (9) & $ 98 $ & (22) & $ 77 $ & (8) & $ 37 $ & (6) & $ 34 $ & (4) & $< 11 $ & \nodata \\
E12-G21         & $ 12 $ & (3) & $< 31 $ & \nodata & $< 18 $ & \nodata & $< 24 $ & \nodata & $ 14 $ & (4) & $< 16 $ & \nodata & $< 14 $ & \nodata & $< 19 $ & \nodata \\
MRK348          & $< 8 $ & \nodata & $< 10 $ & \nodata & $ 8 $ & (2) & $< 12 $ & \nodata & $< 5 $ & \nodata & $< 5 $ & \nodata & $< 7 $ & \nodata & $ 10 $ & (3) \\
NGC424          & $< 3 $ & \nodata & $< 2 $ & \nodata & $< 2 $ & \nodata & $< 7 $ & \nodata & $< 2 $ & \nodata & $< 3 $ & \nodata & $< 2 $ & \nodata & $< 11 $ & \nodata \\
NGC526A         & $ 9 $ & (2) & $ 6 $ & (2) & $ 12 $ & (2) & $ 11 $ & (2) & $ 5 $ & (2) & $ 5 $ & (1) & $ 9 $ & (3) & $< 27 $ & \nodata \\
NGC513          & $< 28 $ & \nodata & $< 44 $ & \nodata & $< 26 $ & \nodata & $< 54 $ & \nodata & $ 38 $ & (9) & $< 28 $ & \nodata & $ 30 $ & (7) & $< 23 $ & \nodata \\
F01475-0740     & $< 35 $ & \nodata & $< 16 $ & \nodata & $ 13 $ & (3) & $< 14 $ & \nodata & $< 6 $ & \nodata & $ 8 $ & (2) & $< 5 $ & \nodata & $< 18 $ & \nodata \\
\enddata
\tablecomments{H$_2$ equivalent widths are given in units of nm. Uncertainties ($1\sigma$) are listed in parentheses. Upper limits are $3\sigma$. The line wavelengths for the S(7) -- S(0) series are 5.512, 6.159, 6.859, 8.004, 9.656, 12.329, 17.063, and 28.171 \m, respectively. }
\end{deluxetable}
\setlength{\tabcolsep}{4pt}

\setlength{\tabcolsep}{2pt} 
\begin{deluxetable}{lrrrrrrrrrrrrrrrrrr}
\rotate
\tabletypesize{\scriptsize}
\tablewidth{0pt}
\tablecaption{Fine Structure Line Fluxes\label{tab:fsstr}}
\tablehead{\colhead{Source ID} & \multicolumn{2}{c}{[OIV]}& \multicolumn{2}{c}{[SiII]}& \multicolumn{2}{c}{[NeII]}& \multicolumn{2}{c}{[NeIII]}& \multicolumn{2}{c}{[NeV]}& \multicolumn{2}{c}{[SIII]}& \multicolumn{2}{c}{[SIII]}& \multicolumn{2}{c}{[SIV]}& \multicolumn{2}{c}{[ArII]}\\
 & \multicolumn{2}{c}{25.91 \m}& \multicolumn{2}{c}{34.815 \m}& \multicolumn{2}{c}{12.813 \m}& \multicolumn{2}{c}{15.555 \m}& \multicolumn{2}{c}{14.322 \m}& \multicolumn{2}{c}{18.713 \m}& \multicolumn{2}{c}{33.48 \m}& \multicolumn{2}{c}{10.511 \m}& \multicolumn{2}{c}{6.985 \m}}
\startdata
MRK335          & $ 32 $ & (7) & $< 80 $ & \nodata & $ 24 $ & (5) & $ 44 $ & (9) & $< 19 $ & \nodata & $< 32 $ & \nodata & $< 87 $ & \nodata & $ 27 $ & (8) & $< 70 $ & \nodata \\
MRK938          & $< 126 $ & \nodata & $< 796 $ & \nodata & $ 563 $ & (25) & $< 82 $ & \nodata & $< 56 $ & \nodata & $ 52 $ & (15) & $< 895 $ & \nodata & $< 55 $ & \nodata & $ 209 $ & (30) \\
E12-G21         & $ 202 $ & (13) & $ 169 $ & (52) & $ 140 $ & (19) & $ 61 $ & (8) & $ 52 $ & (9) & $ 90 $ & (9) & $< 174 $ & \nodata & $ 47 $ & (12) & $< 69 $ & \nodata \\
MRK348          & $ 179 $ & (14) & $ 131 $ & (35) & $ 131 $ & (10) & $ 178 $ & (16) & $ 54 $ & (11) & $ 42 $ & (12) & $< 110 $ & \nodata & $ 62 $ & (9) & $ 69 $ & (14) \\
NGC424          & $ 145 $ & (29) & $< 244 $ & \nodata & $ 73 $ & (22) & $ 136 $ & (30) & $< 58 $ & \nodata & $< 53 $ & \nodata & $< 244 $ & \nodata & $< 54 $ & \nodata & $< 68 $ & \nodata \\
NGC526A         & $ 203 $ & (10) & $< 155 $ & \nodata & $ 51 $ & (11) & $ 135 $ & (13) & $< 25 $ & \nodata & $ 56 $ & (10) & $ 66 $ & (19) & $ 61 $ & (12) & $ 57 $ & (10) \\
NGC513          & $ 110 $ & (8) & $ 308 $ & (36) & $ 237 $ & (15) & $ 93 $ & (9) & $ 24 $ & (8) & $ 100 $ & (8) & $ 217 $ & (38) & $< 48 $ & \nodata & $< 94 $ & \nodata \\
F01475-0740     & $< 65 $ & \nodata & $< 182 $ & \nodata & $ 157 $ & (9) & $ 96 $ & (10) & $ 21 $ & (6) & $ 47 $ & (11) & $< 185 $ & \nodata & $< 30 $ & \nodata & $ 62 $ & (19) \\
\enddata
\tablecomments{Integrated fluxes are given in units of $10^{-18}$~W~m\mtwo. Uncertainties ($1\sigma$) are listed in parentheses. Upper limits are $3\sigma$.  }
\end{deluxetable}
\setlength{\tabcolsep}{4pt}

\setlength{\tabcolsep}{2pt} 
\begin{deluxetable}{lrrrrrrrrrrrrrrrrrr}
\rotate
\tabletypesize{\scriptsize}
\tablewidth{0pt}
\tablecaption{Fine Structure Line EQWs\label{tab:fseqw}}
\tablehead{\colhead{Source ID} & \multicolumn{2}{c}{[OIV]}& \multicolumn{2}{c}{[SiII]}& \multicolumn{2}{c}{[NeII]}& \multicolumn{2}{c}{[NeIII]}& \multicolumn{2}{c}{[NeV]}& \multicolumn{2}{c}{[SIII]}& \multicolumn{2}{c}{[SIII]}& \multicolumn{2}{c}{[SIV]}& \multicolumn{2}{c}{[ArII]}\\
 & \multicolumn{2}{c}{25.91 \m}& \multicolumn{2}{c}{34.815 \m}& \multicolumn{2}{c}{12.813 \m}& \multicolumn{2}{c}{15.555 \m}& \multicolumn{2}{c}{14.322 \m}& \multicolumn{2}{c}{18.713 \m}& \multicolumn{2}{c}{33.48 \m}& \multicolumn{2}{c}{10.511 \m}& \multicolumn{2}{c}{6.985 \m}}
\startdata
MRK335          & $ 22 $ & (5) & $< 101 $ & \nodata & $ 7 $ & (2) & $ 15 $ & (3) & $< 6 $ & \nodata & $< 12 $ & \nodata & $< 102 $ & \nodata & $ 5 $ & (2) & $< 9 $ & \nodata \\
MRK938          & $< 9 $ & \nodata & $< 46 $ & \nodata & $ 125 $ & (6) & $< 13 $ & \nodata & $< 10 $ & \nodata & $ 8 $ & (2) & $< 53 $ & \nodata & $< 30 $ & \nodata & $ 83 $ & (12) \\
E12-G21         & $ 170 $ & (11) & $ 158 $ & (49) & $ 68 $ & (9) & $ 36 $ & (5) & $ 28 $ & (5) & $ 64 $ & (6) & $< 163 $ & \nodata & $ 18 $ & (5) & $< 18 $ & \nodata \\
MRK348          & $ 65 $ & (5) & $ 83 $ & (22) & $ 24 $ & (2) & $ 33 $ & (3) & $ 10 $ & (2) & $ 8 $ & (3) & $< 64 $ & \nodata & $ 12 $ & (2) & $ 9 $ & (2) \\
NGC424          & $ 23 $ & (5) & $< 70 $ & \nodata & $ 4 $ & (1) & $ 8 $ & (2) & $< 3 $ & \nodata & $< 4 $ & \nodata & $< 65 $ & \nodata & $< 2 $ & \nodata & $< 2 $ & \nodata \\
NGC526A         & $ 148 $ & (7) & $< 290 $ & \nodata & $ 11 $ & (3) & $ 35 $ & (3) & $< 6 $ & \nodata & $ 18 $ & (3) & $ 108 $ & (31) & $ 10 $ & (2) & $ 9 $ & (2) \\
NGC513          & $ 94 $ & (7) & $ 256 $ & (30) & $ 149 $ & (9) & $ 66 $ & (7) & $ 16 $ & (5) & $ 75 $ & (6) & $ 184 $ & (32) & $< 24 $ & \nodata & $< 33 $ & \nodata \\
F01475-0740     & $< 23 $ & \nodata & $< 100 $ & \nodata & $ 41 $ & (2) & $ 24 $ & (3) & $ 5 $ & (2) & $ 11 $ & (3) & $< 96 $ & \nodata & $< 6 $ & \nodata & $ 17 $ & (5) \\
\enddata
\tablecomments{Equivalent widths are given in units of nm. Uncertainties ($1\sigma$) are listed in parentheses. Upper limits are $3\sigma$.  Uncertainties ($1\sigma$) are listed in parentheses. Upper limits are $3\sigma$.  }
\end{deluxetable}
\setlength{\tabcolsep}{4pt}

\begin{deluxetable}{llrrrr}
\tabletypesize{\scriptsize}
\tablewidth{0pt}
\tablecaption{Dust Opacities and MIPS Aperture Corrections determined
  by PAHFIT\label{tab:tauapcor}}
\tablehead{
\colhead{Source ID} & \colhead{Model} & \multicolumn{2}{c}{$\tau_{10}$}  &
\multicolumn{2}{c}{Ap. Corr.}\\
\colhead{(1)} & \colhead{(2)} & \multicolumn{2}{c}{(3)}  &
\multicolumn{2}{c}{(4)}
}
\startdata
MRK335 & Screen & 0.10 & (0.07) & 1.00 & \nodata\\
MRK938 & Screen & 0.91 & (0.03) & 1.27 & (0.02)\\
E12-G21 & Mixed & 1.21 & (0.26) & 1.00 & \nodata\\
MRK348 & Screen & 0.22 & (0.08) & 1.66 & (0.06)\\
NGC424 & \nodata & 0.00 & \nodata & 1.14 & (0.07)\\
NGC526A & \nodata & 0.00 & \nodata & 1.81 & \nodata\\
NGC513 & Screen & 0.24 & (0.02) & 1.23 & (0.02)\\
F01475-0740 & Screen & 0.09 & (0.10) & 1.00 & \nodata\\
\enddata
\tablecomments{(1) Source ID. (2) Dust extinction model: {\em Screen}
  models place all of the obscuring dust between the observer and the
  infrared continuum sources, and {\em Mixed} models assume that the
  obscuring dust is uniformly mixed with the infrared continuum
  sources. (3) Opacity of the obscuring dust component at 10~\m; this
  opacity affects all of the continuum model components except for the
  optically-thin hot dust component.  (4) The aperture correction for
  the MIPS-SED data; the MIPS-SED data need to be divided by this
  correction factor to reconcile the model fit with the smaller IRS
  and IRAC apertures. The correction factor was restricted to the
  range 1 -- 1.81 during the fitting procedure.}
\end{deluxetable}

\begin{deluxetable}{lrrrr}
\tablewidth{0pt}
\tabletypesize{\scriptsize}
\tablecaption{Silicate Strengths\label{tab:silicates}}
\tablehead{
Source ID &  \multicolumn{2}{c}{$S_{10}$} &  \multicolumn{2}{c}{$S_{18}$}
}
\startdata
MRK335          & $ 0.295$ & (0.028)& $ 0.089$ & (0.009) \\
MRK938          & $-0.536$ & (0.046)& $-0.268$ & (0.009) \\
E12-G21         & $ 0.036$ & (0.050)& $-0.001$ & (0.019) \\
MRK348          & $-0.120$ & (0.022)& $ 0.079$ & (0.011) \\
NGC424          & $ 0.111$ & (0.018)& $ 0.085$ & (0.011) \\
NGC526A         & $ 0.252$ & (0.031)& $ 0.098$ & (0.011) \\
NGC513          & $ 0.030$ & (0.067)& $ 0.089$ & (0.095) \\
F01475-0740     & $ 0.143$ & (0.017)& $ 0.163$ & (0.007) \\
\enddata
\tablecomments{Sil 10 \m\ and 18 \m\ strengths. Sil absorption is indicated by negative strengths, emission, positive. Formal statistical uncertainties are given in parentheses. }
\end{deluxetable}

\begin{deluxetable}{lrrrr}
\tablewidth{0pt}
\tabletypesize{\scriptsize}
\tablecaption{Infrared Continuum Spectral Indices\label{tab:spectralindices}}
\tablehead{
Source ID &  \multicolumn{2}{c}{$\alpha$(20--30 \m)} &  \multicolumn{2}{c}{$\alpha$(55-90 \m)}
}
\startdata
MRK335 & $-$1.75 & (0.03) & $-$1.49 & (0.29) \\
MRK938 & 1.58 & (0.02) & $-$2.03 & (0.03) \\
E12-G21 & $-$0.62 & (0.02) & $-$0.59 & (0.10) \\
MRK348 & $-$1.87 & (0.01) & $-$2.62 & (0.23) \\
NGC424 & $-$2.23 & (0.02) & $-$2.60 & (0.11) \\
NGC526A & $-$2.91 & (0.03) & $-$1.50 & (0.44) \\
NGC513 & $-$0.40 & (0.03) & $-$0.72 & (0.06) \\
F01475-0740 & $-$1.36 & (0.03) & $-$2.23 & (0.19) \\
\enddata
\tablecomments{Continuum spectral indices over the wavelength ranges 20--30 \m\ and 55-90 \m. Formal statistical uncertainties are provided in parentheses.The convention is $F_{\lambda} \propto \lambda^{\alpha}$. }
\end{deluxetable}

\clearpage

\epsscale{0.8}
\clearpage

\epsscale{0.80}
\plotone{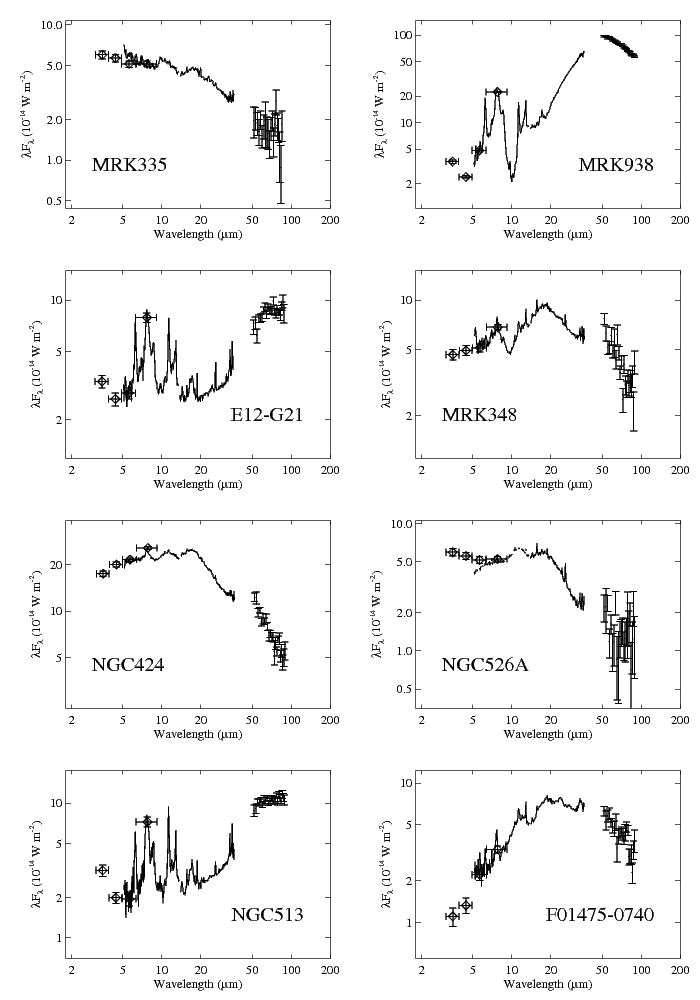}

\begin{figure}
\plotone{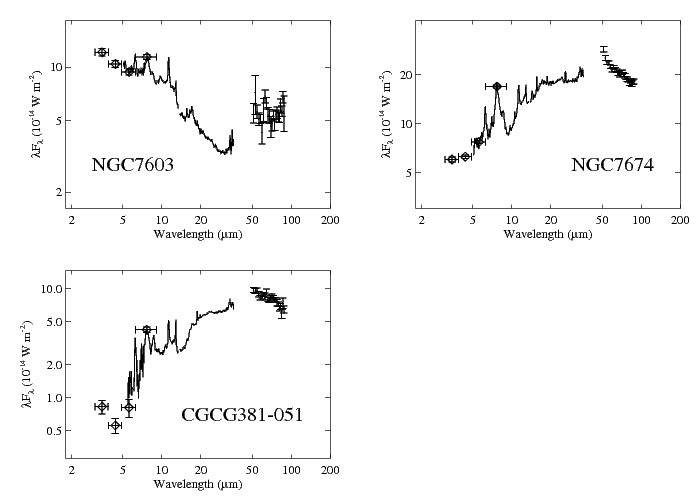}
\caption{{\em Spitzer} spectrophotometry of the 12 \m\ AGN sample. The
  SEDs are plotted as $\lambda F_{\lambda}$ vs. rest $\lambda$ and in
  order of increasing RA. IRAC data are plotted as diamonds with
  horizontal errorbars representing the camera bandwidth and vertical
  errorbars indicating a combination of statistical uncertainties,
  uncertainty of the extended flux correction, and uncertainty of the
  color corrections. IRS data are traced by the solid line; each camera
  and order is plotted independently, hence there are gaps appearing
  around 15 \m, which separates the SL and LL spectrographs. Scaled,
  staring-mode IRS data are indicated by dash-dot lines. MIPS SED data
  are show as vertical error bars, which represent statistical
  uncertainties of the MIPS-SED photometry, but do not include the
  systematic 10-15\% uncertainty associated with the flux calibration.
  The IRAC and IRS data were extracted from 20\arcsec\ diameter
  synthetic apertures, and the MIPS-SED data were extracted from
  30\arcsec\ aperture along a 20\arcsec\ wide slit.}\label{fig:allseds}
\end{figure}

Figure~\ref{fig:allseds}
\clearpage
\plotone{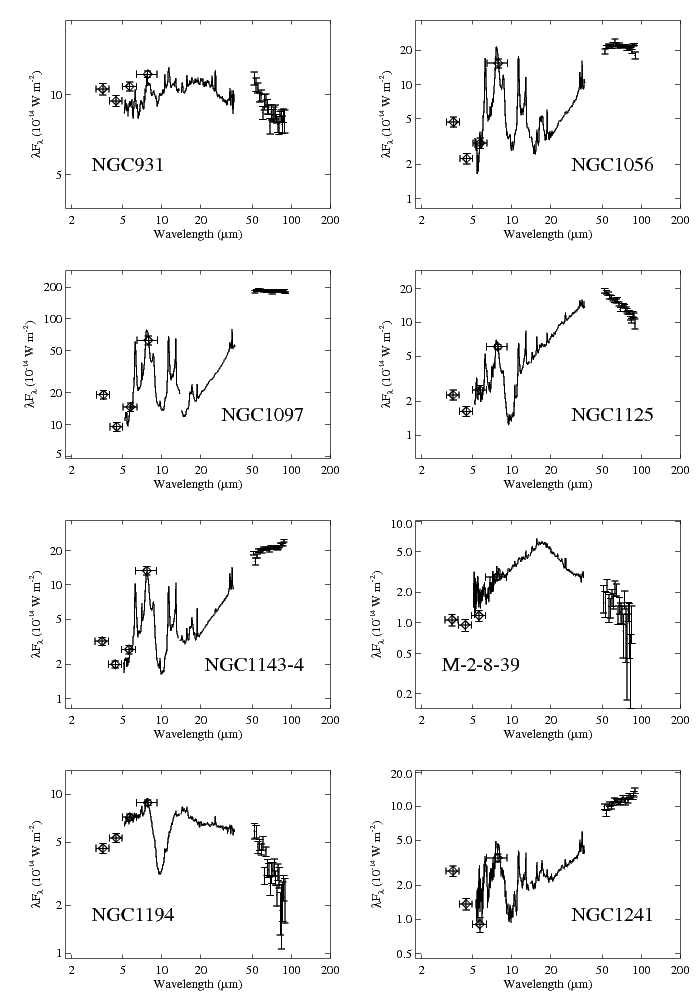}

Figure~\ref{fig:allseds}
\clearpage
\plotone{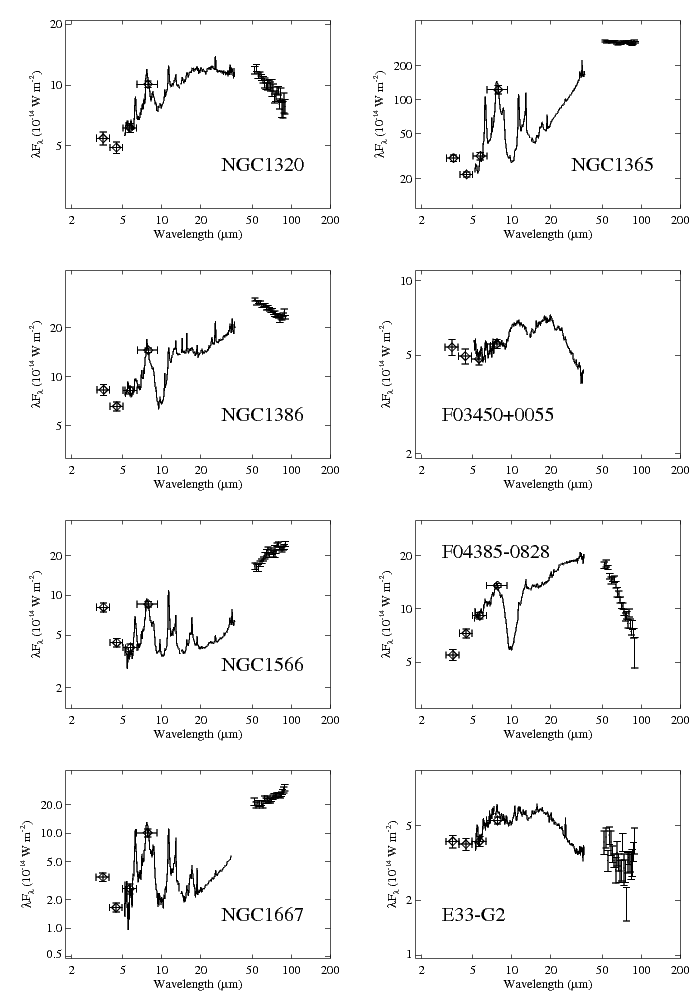}

Figure~\ref{fig:allseds}
\clearpage
\plotone{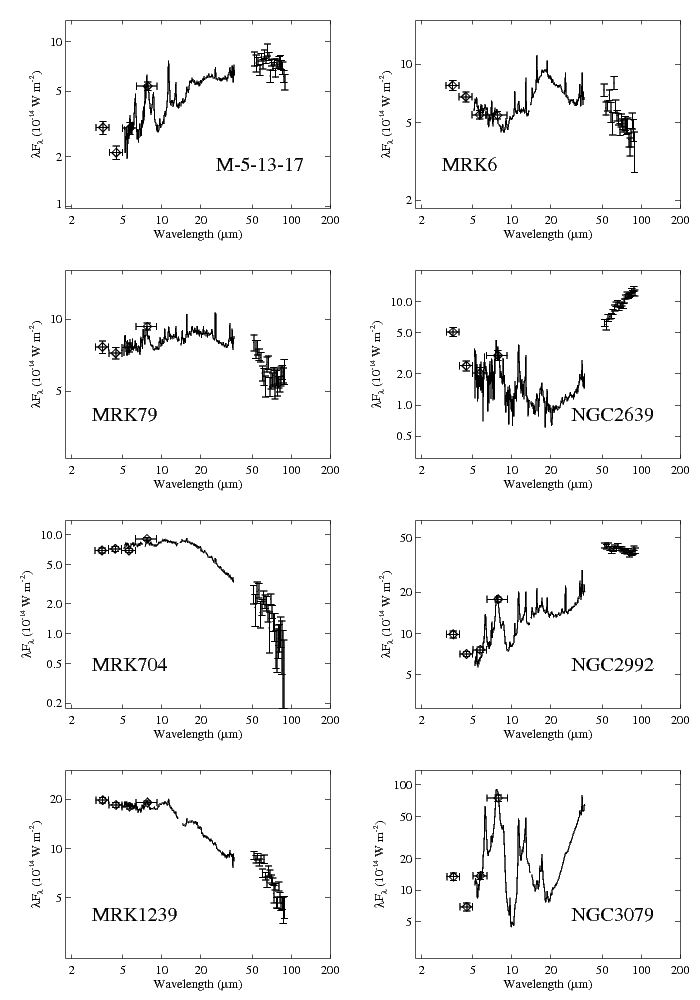}

Figure~\ref{fig:allseds}
\clearpage
\plotone{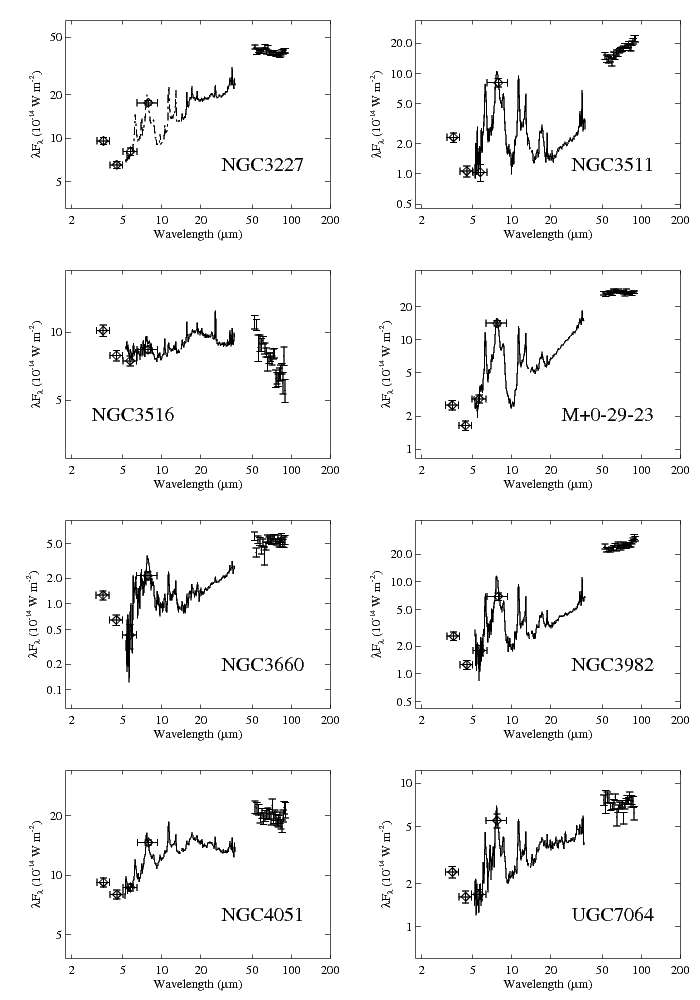}

Figure~\ref{fig:allseds}
\clearpage
\plotone{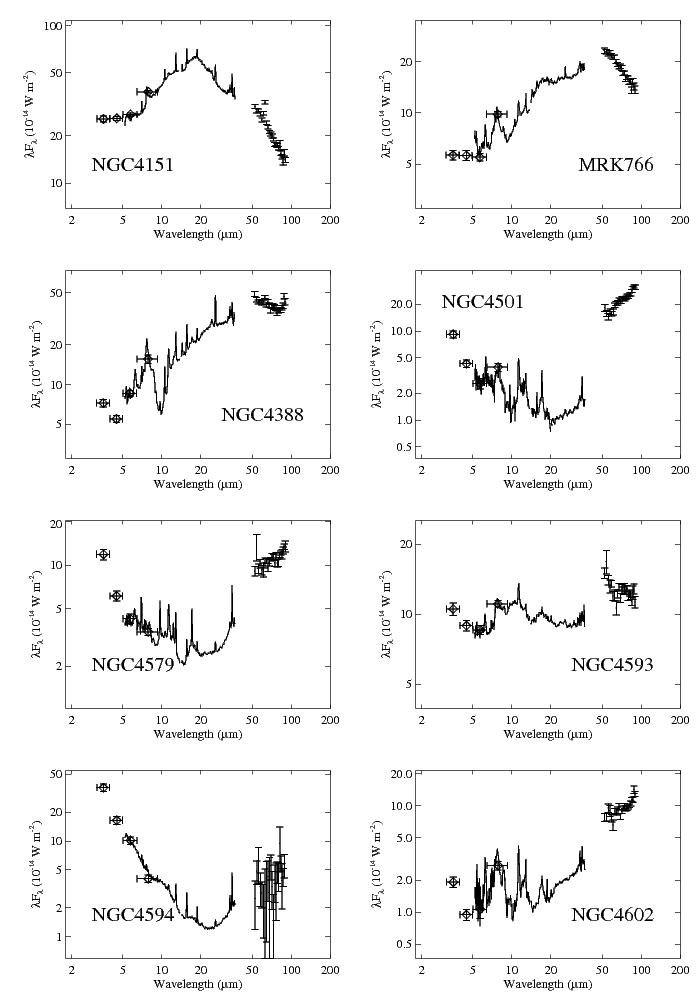}

Figure~\ref{fig:allseds}
\clearpage
\plotone{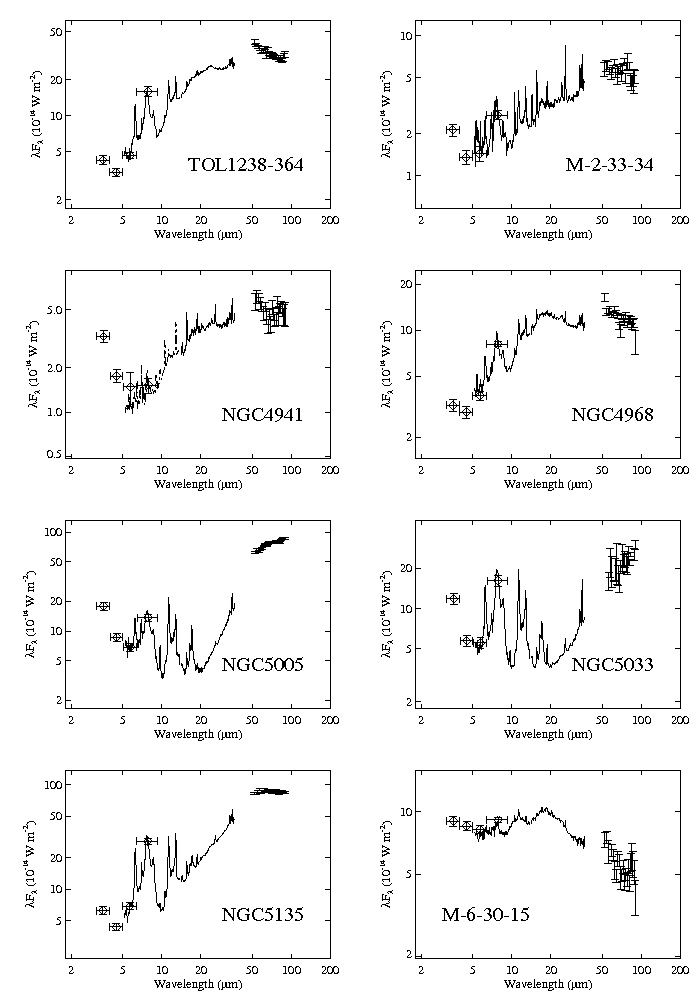}

Figure~\ref{fig:allseds}
\clearpage
\plotone{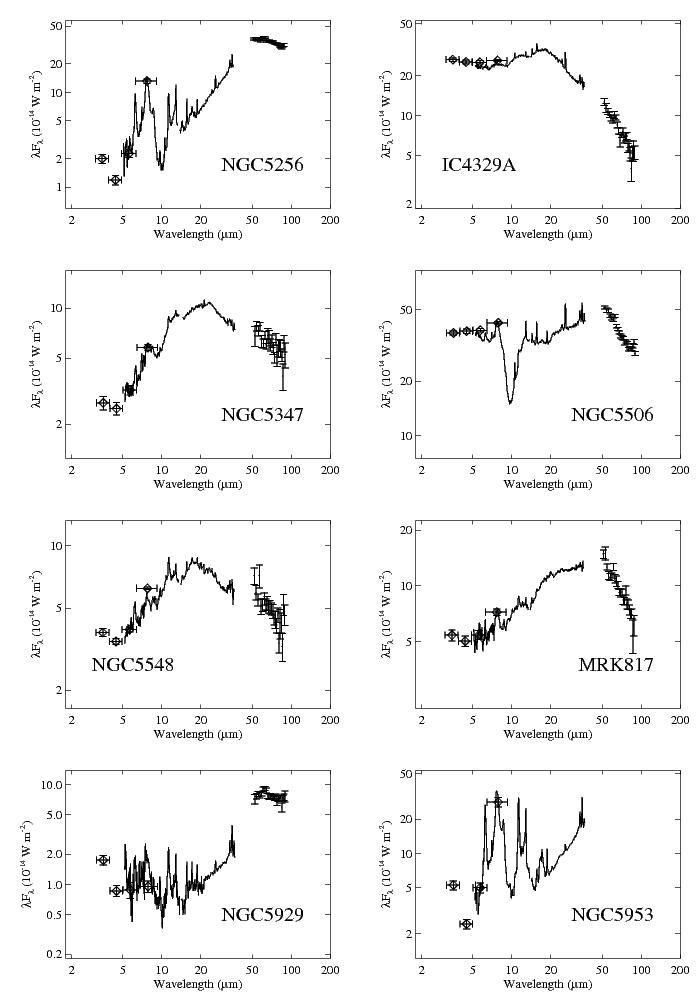}

Figure~\ref{fig:allseds}
\clearpage
\plotone{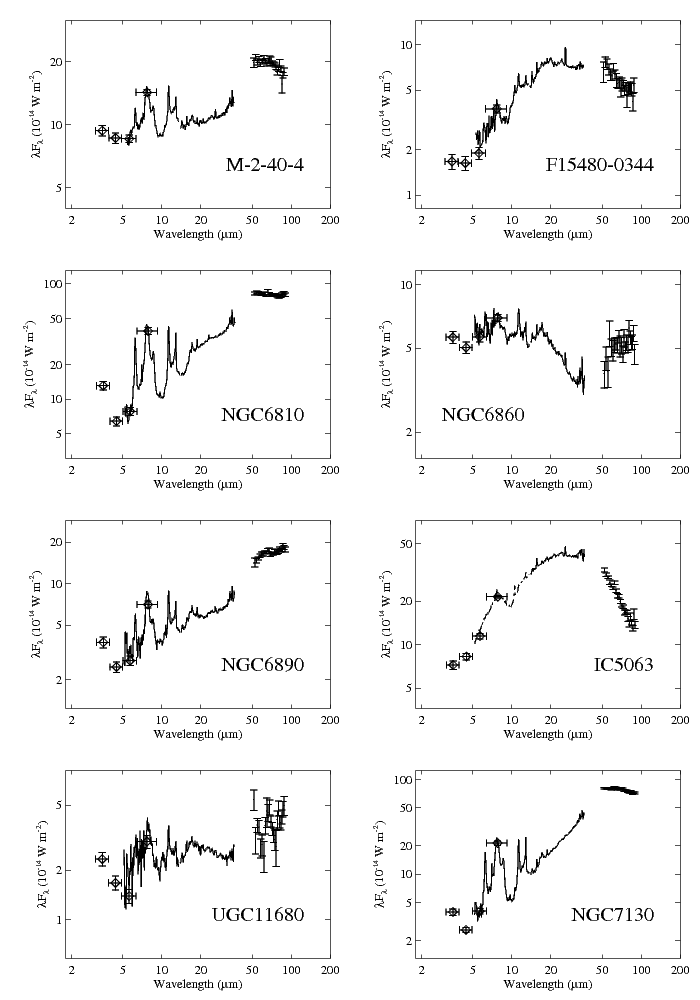}

Figure~\ref{fig:allseds}
\clearpage
\plotone{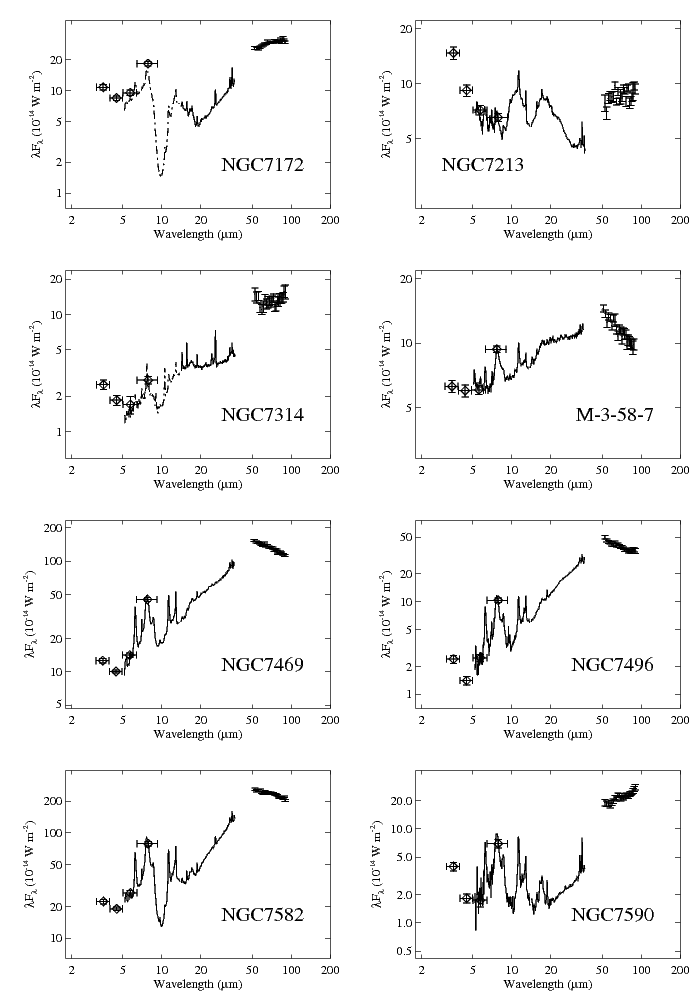}

Figure~\ref{fig:allseds}
\clearpage

\epsscale{1.00}

\clearpage

\epsscale{1.00}

\clearpage

\epsscale{1.00}

\epsscale{1.00}

\epsscale{1.00}

\clearpage

\end{document}